\documentclass[11pt]{article}

\newlength{\medpaperheight}     
\newlength{\medpaperwidth}      
\newlength{\medtextheight}      
\newlength{\medtextwidth}       
\newlength{\medtopmargin}       
\newlength{\medoddsidemargin}   

\newcommand{\diag}{{\rm diag}}

\newcommand{\gam}{\gamma}

\newcommand{\om}{\omega}
\newcommand{\lam}{\lambda}
\newcommand{\bet}{\beta}

\newcommand{\del}{\delta}
\newcommand{\sig}{\sigma}

\newcommand{\Om}{\Omega}

\newcommand{\norm}[1]{ \left\| #1 \right\| }

\newcommand{\delh}{\hat{\del}}
\newcommand{\rhoh}{\hat{\rho}}

\newcommand{\eg}{\emph{e.g.}}
\newcommand{\ie}{\emph{i.e.}}

\newcommand{\bquem}{\begin{quote}\begin{em}}
\newcommand{\equem}{\end{em}\end{quote}}

\newcommand{\blist}{\begin{description}}
\newcommand{\elist}{\end{description}}

\newcommand{\bquote}{\begin{quote}}
\newcommand{\equote}{\end{quote}}

\newcommand{\ben}{\begin{enumerate}}
\newcommand{\een}{\end{enumerate}}

\newcommand{\bit}{\begin{itemize}}
\newcommand{\eit}{\end{itemize}}

\newcommand{\bea}{\begin{array}}
\newcommand{\eea}{\end{array}}

\newcommand{\bds}{\begin{displaystyle}}
\newcommand{\eds}{\end{displaystyle}}

\newcommand{\Rbf}{{\mathbf R}}
\newcommand{\Cbf}{{\mathbf C}}

\newcommand{\Dbf}{{\mathbf D}}

\newcommand{\ds}{\displaystyle}

\newcommand{\refeq}[1]{(\ref{eq:#1})}

\newcommand{\refsec}[1]{\ref{sec:#1}}

\newcommand{\set}[2]{ \left\{ \,#1\, \left| \,#2\, \right.\right\} }

\newcommand{\seq}[1]{ \left\{ #1 \right\} }

\newcommand{\opt}{{\rm opt}}

\newcommand{\mathbox}[1]{
\fbox{$\ds #1 $}
}

\makeatletter
\def\beq{\@ifnextchar 
[{\@tempswatrue\@beq}{\@tempswafalse\@beq[]}}
\def\@beq[#1]{\begin{equation}\edef\@tmparg{#1}\ifx\@tmparg\@e
mpty \else
	\label{#1}\fi}
\newcommand{\eeq}{\end{equation}}
\makeatother 
\newcommand{\beqaa}{\begin{eqnarray*}}
\newcommand{\eeqaa}{\end{eqnarray*}}

\newcommand{\beqa}{\begin{eqnarray}}
\newcommand{\eeqa}{\end{eqnarray}}

\newcommand{\bc}{\begin{center}}
\newcommand{\ec}{\end{center}}

\usepackage{epsfig}
\usepackage{times}
\usepackage{amsmath,amssymb}

\usepackage{psfrag}
\newcommand{\figsfile}[1]
{\epsfig{file=#1}}
\renewcommand{\diag}{{\bf diag}}
\newcommand{\apost}{{\em a posteriori\ }}

\newcommand{\sigmax}{\sig_{\max}}

\newcommand{\incl}{{\rm incl}}
\newcommand{\pb}{\bar{p}}

\newcommand{\sh}{\hat{s}}
\renewcommand{\dh}{\hat{d}}

\newcommand{\pdyn}{p_\dyn}
\newcommand{\dyn}{{\rm dyn}}

\newcommand{\normav}[1]{\norm{#1}_{\rm avg}}
\newcommand{\normwc}[1]{\norm{#1}_{\rm wc}}

\newcommand{\pin}{p_{\rm in}}
\newcommand{\pout}{p_{\rm out}}
\newcommand{\pinout}{p_{{\rm in} | {\rm out}}}
\newcommand{\poutin}{p_{{\rm out} | {\rm in}}}
\newcommand{\pjoint}{p_{\rm joint}}
\newcommand{\pincl}{p_{\rm incl}}

\renewcommand{\seq}[1]{\{ #1 \}}

\newcommand{\post}{{\rm post}}
\newcommand{\ppost}{p_\post}
\newcommand{\cond}{{\rm cond}}

\newcommand{\joint}{{\rm joint}}
\newcommand{\noisy}{{\rm noisy}}

\newcommand{\mh}{{\hat{m}}}

\newcommand{\rank}{{\bf rank}}

\newcommand{\qsys}{Q}

\newcommand{\Cbfnn}{{\bf C}^{n\times n}}

\newcommand{\prob}[1]{{\bf Prob}\left\{#1\right\}}
\newcommand{\trace}{{\bf Tr}}

\begin{document}

\title{
Quantum State Detector Design: 
\\
Optimal Worst-Case \apost Performance
%via Convex Optimization
%
\footnote{Kosut, Walmsley, Rabitz supported by the DARPA QUIST Program.}
}
\author{
Robert L. Kosut\thanks{
SC Solutions, Sunnyvale, CA, USA,
{\tt kosut@scsolutions.com}
}
\and
Ian Walmsley\thanks{
Oxford University,
Oxford, UK,
{\tt walmsley@physics.ox.ac.uk}
}
\and
Yonina Eldar\thanks{
Technion,
Haifa, Israel,
{\tt yonina@ee.technion.ac.il}
}
\and
Herschel Rabitz\thanks{
Princeton University,
Princeton, NJ,
{\tt hrabitz@princeton.edu}
}
}

%\date{v07d \today}
\date{March 21, 2004}

\maketitle
%%%%%%%%%%%%%%%%

\begin{abstract}

The problem addressed is to design a detector which is maximally
sensitive to specific quantum states. Here we concentrate on quantum
state detection using the {\em worst-case} \apost {\em probability of
detection} as the design criterion. This objective is equivalent to
asking the question: if the detector declares that a specific state is
present, what is the probability of that state actually being present?
We show that maximizing this worst-case probability (maximizing the
smallest possible value of this probability) is a {\em quasiconvex
optimization} over the matrices of the POVM (positive operator valued
measure) which characterize the measurement apparatus. We also show
that with a given POVM, the optimization is quasiconvex in the matrix
which characterizes the Kraus operator sum representation (OSR) in a
fixed basis.  We use Lagrange Duality Theory to establish the
optimality conditions for both deterministic and randomized
detection. We also examine the special case of detecting a single pure
state. Numerical aspects of using convex optimization for quantum
state detection are also discussed.

\end{abstract}

%%%%%%%%%%%%%%%%%%%%%%%%%%%%%%%%%
%%%%%%%%%%%%%%%%%%%%%%%%%%%%%%%%
\newpage
\section{Introduction}

\paragraph{why}

Information is extracted from a quantum system by measurement.  The
most information that it is possible to extract from the quantum
system is given by its state, specified by a density operator, and it
is impossible to determine this from a single measurement. The problem
of detecting information stored in the state of a quantum system is
therefore a fundamental problem in quantum information theory.  The
nature of this problem is essentially the design of measurements such
that they yield the optimum information for the specified
purpose. That is, the construction of matrices representing positive
operator valued measures (POVMs) which give the best performance
against a given set of criteria, subject to constraints reflecting the
underlying properties of the quantum mechanics, or costs associated
with the implementation of certain operations.

The emergence of quantum information processing has raised important
new issues, and made more urgent the development of tools for the
design of quantum measurements. In this paper we present a general
formalism that enables this design across a wide range of
applications. In particular we show that the problem may be cast in
the form of a convex optimization over the possible POVMs, and that
this allows powerful numerical tools to identify the globally optimal
measurements to achieve the desired objective.  Such optimizations are
useful even if they turn out to be difficult to implement in the
laboratory, since they provide a benchmark for the performance of
experimentally feasible measurements.  

The objective of a measurement in quantum information theory depends
on the way in which information is encoded into the quantum system to
begin with, and this is in turn, depends on the application. In
quantum cryptography, for example, the information is encoded by the
sender choosing randomly between two non-orthogonal bases, both of
which can encode a single bit.  The ability of an eavesdropper, who is
in principle unable to influence the choice of preparation and
measurement bases chosen by the sender and receiver of this
information, to extract information from the transmitted quantum bits,
depends on her ability to determine which of four non-orthogonal
states were sent.  In a quantum information processor, the information
in the register at the end of a computation often resides in
orthogonal states, and the goal of the measurement is simply to read
out the register by distinguishing among the sets of such states.
However, the operation of such a processor may itself depends on
measurements. For example, quantum error correction protocols require
the measurement of an ancilla to preserve the quantum state of the
register itself.  In another example, the {\em cluster computing
model} \cite{cluster} and the {\em linear optical quantum computer}
\cite{loqc}, both rely on measurements of ancillary qubits for the
operation of the logic gates themselves.

In these examples of conditional state preparation, it is vital that
there is a high degree of correlation between the outcome of the
measurement and the quantum state prepared in the register by the
measurement. The structure of the measurement should therefore be such
that this correlation is maximized. Thus it is vital to consider the
case when the detectors have noise, and to develop strategies for
optimizing the measurement in the presence of this and inherent
inefficiencies.  Despite the fundamental inability to determine the
quantum state of a system from a single measurement, it is sometimes
useful to make such a determination from a set of independent
measurements on identically prepared systems. This procedure is called
{\em quantum state tomography.} Similarly, one may characterize the
action of a quantum operation by determining its effect on a known
input quantum state from a determination of the output state.  In this
application, the central questions are: how many measurements are
needed to determine the state to within a given precision, and how
should these measurements be constructed? That is, it is essentially a
problem of {\em experiment design}. Optimal experiment for quantum
state tomography and quantum process tomography is considered in
\cite{KosutWEH:04}.

In this work, we concentrate on the problem of {\em quantum state
detection}. That is, the design of POVMs that can determine whether or
not a particular component was present in the input state to the
detector. The problem is thus equivalent to the design of a quantum
channel that optimally transforms the input state distribution
(assumed to be given, and including non-orthogonal states) to the
output measurement outcomes. The channel may be lossy, and may
introduce noise, and thus there may be latency in the measurement, in
which certain outcomes are ambiguous.

\paragraph{previous work}

Several approaches have emerged for distinguishing between a
collection of non-orthogonal quantum states.  An accessible review can
be found in the article by Chefles \cite{C00}.  In one approach,
called quantum hypothesis testing, a measurement is designed to
minimize the probability of a detection error
\cite{H73,H76,YKL75,EMV02,CBH89,OBH96,BKMO97,EF01,EMV02s,E03l}.
Necessary and sufficient conditions for an optimum measurement
maximizing the probability of correct detection have been developed in
\cite{EMV02} using a semidefinite programming approach, and earlier in
\cite{H73} (the drawback of this approach is that it does not readily
lend itself to efficient computational algorithms). Closed-form
analytical expressions for the optimal measurement have been derived
for several special cases
\cite{H76,CBH89,OBH96,BKMO97,EF01,EMV02s}. Iterative procedures
maximizing the probability of correct detection have also been
developed for cases in which the optimal measurement cannot be found
explicitly \cite{H82,EMV02}.  A specific design for achieving the
optimal discrimination between non-orthogonal coherent states has been
given in \cite{K99}, and for non-orthogonal polarization states of a
single photon by \cite{Barnett:97}. Optimal discrimination amongst more than
two non-orthogonal states has also been analyzed \cite{P00} and
demonstrated experimentally \cite{C01}.

A more recent approach, referred to as unambiguous detection
\cite{I87,D88,P88,JS95,PT98,C98,CB98,E02,E03p,ESB03}, is to design a
measurement that with a certain probability returns an inconclusive
result, but such that if the measurement returns an answer, then the
answer is correct with probability $1$. Chefles \cite{C98} showed that
a necessary and sufficient condition for the existence of unambiguous
measurements for distinguishing between a collection of {\em pure}
quantum states is that the states are linearly independent. Necessary
and sufficient conditions on the optimal measurement minimizing the
probability of an inconclusive result for pure states were derived in
\cite{E02}. The optimal measurement when distinguishing between a
broad class of symmetric pure-state sets was also considered in
\cite{E02}. The problem of unambiguous detection between {\em mixed}
state ensembles was first considered in \cite{RST03}.  Necessary and
sufficient optimality conditions for unambiguous mixed state detection
were developed in \cite{ESB03}.

Experimental configurations may not allow the ideal measurements to be
made, and thus the performance of feasible apparatuses have been
analyzed. For example, an apparatus for the unambiguous discrimination
between two orthogonal states of a single photon using homodyne
detection, rather than photon counting, which has higher losses and
more noise, has been examined in \cite{GriceW:95} and \cite{B00}.  An
experimental implementation of the process for discriminating
unambiguously between two non-orthogonal states of polarization of a
single photon has been demonstrated in \cite{Huttner:96}.

An interesting alternative approach for distinguishing between a
collection of quantum states, which is a combination of the previous
two approaches, is to allow for a certain probability of an
inconclusive result, and then maximize the probability of correct
detection \cite{E03p,ZLG99,FJ02}.

\paragraph{what's new here} 

Prior work has considered optimal detector design only for an average
measure of the probability of detection, such as the {\em average
joint probability of detection}. For example, in \cite{EMV02,E02} it
is shown that using this criterion, detector design can be formulated
as a convex optimization over the matrices in the POVM, specifically a
{\em semidefinite program} (SDP). Here we concentrate on quantum state
detection using the {\em worst-case} \apost {\em probability of
detection} as the design criterion. This objective is equivalent to
asking the question: if the detector declares that a specific state is
present, what is the probability of that state actually being present?
We show that maximizing the smallest possible value of this
probability is a {\em quasiconvex optimization} over the POVM
matrices, or over the Krause operator-sum-representation (OSR) in a
fixed basis.  Issues relating to conditions of optimality and
numerical aspects of convex optimization of state detection are also
discussed.

We will show that many of the standard measures of detector
performance (including those previously considered) are also convex
functions of the detector design parameters.  In addition, we will see
that the design parameters, either POVM or OSR, are in a convex set.
As a result we can cast a number of detector design problems as a
convex optimization.  Details and underlying theory about convex
optimization are in the text by Boyd and Vandenberghe
\cite{BoydV:04}. As stated there, the great advantage of convex
optimization is a globally optimal solution can be found efficiently
and reliably, and perhaps most importantly, can be computed to within
any desired accuracy using an {\em interior-point method}.

Another advantage to being able to obtain a globally optimal solution
is that the resulting performance can be used as a benchmark against
which the initial detector design can be compared. If the optimal
performance is significantly better, then there is compelling reason
to try and implement the optimal solution or to try and modify the
initial design in the ``direction'' of the optimal solution, if that
is clear from the physical implementation.

In a few instances we use Lagrange Duality Theory to derive formulas
for direct calculation of the optimal objective value and the
associated POVM matrices. These calculations only involve singular
value decomposition of the problem data.

%%%%%%%%%%%%%%%%%%%%%%%%%%%%%%%%
\section{Problem formulation}

%%%%%%%%%%%%%%%%%%%%%%%%%%%%%%%%%%%%%%%%
\subsection{Detector}
\label{sec:detector}

\begin{figure}[h]
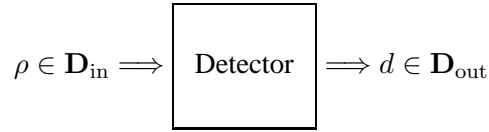

\[
\rho\in\Dbf_{\rm in}
\Longrightarrow
\mathbox{
\bea{c}
\\
\mbox{Detector}
\\
\mbox{}
\eea
}
\Longrightarrow
d \in\Dbf_{\rm out}
\]
\caption{Quantum state detector}
\label{fig:detector}
\end{figure}

A {\em quantum state detector} is considered here as an input/output
device mapping a state (density matrix) $\rho\in\Cbfnn$ at the input
into one of a number of discrete outcomes at the output as illustrated
in Figure \ref{fig:detector}.

\noindent
Specifically, the input state is drawn randomly from
\beq[eq:omin]
\Dbf_{\rm in}
=
\set{\rho_i\in\Cbfnn,\ 0 \leq p_i \leq 1}{i=1,\ldots,m}
\eeq
where $p_i$ is the occurence probability of $\rho_i$, that is,
\beq[eq:pin]
p_i=\prob{\rho=\rho_i},\; i=1,\ldots,m
\eeq
The set of detector outcomes is, 
\beq[eq:out]
\Dbf_{\rm out}
=
\set{i}{i=1,\ldots,m} 
\eeq
The problem addressed is to design the detector to be able to
determine the presence of some or all of the specified set of input
states given knowledge of the input set $\Dbf_{\rm in}$ and the
associated occurrence probabilities. Although the principal focus is
on an equal number of state inputs and detector outcomes, this is not
always the case , \eg, noisy measurements can result in unequal inputs
and outcomes as briefly discussed in Section \refsec{noisy
meas}.

%%%%%%%%%%%%%%%%%%%%%%%%%%%%%%%%%%%%%%%%
\subsection{Performance probabilities}
\label{sec:probs}

Detector performance is usually assessed by examination of one or more
of the following probability matrices:
\beq[eq:probmats]
\bea{llll}
\mbox{joint probability matrix}
&
\pjoint(i,j)
&=&
\prob{\mbox{detect $i$ AND input $j$}}
\\&&&\\
\mbox{conditional probability matrix}
&
\poutin(i|j)
&=&
\prob{\mbox{detect $i$ GIVEN input $j$}}
\\&&&\\
\mbox{\apost probability matrix}
&
\pinout(j|i)
&=&
\prob{\mbox{input $j$ GIVEN detect $i$}}
\eea
\eeq
As shown in any standard text \cite{GonickS:93} these probabilities
are related as follows:
\beq[eq:allprob]
\bea{rcl}
\pjoint(i,j)
&=&
\poutin(i|j) p_j 
=
\pinout(j|i) \pout(i)
\\
\pout(i) 
&=& 
\sum_{j=1}^m\ \poutin(i|j) p_j 
\eea
\eeq
Without loss of generality we can order the input and output events so
that detector event 1 corresponds to input event 1, detector event 2
to input event 2, and so on. With this ordering, detector performance
can be assessed by the {\em error probabilities}:
\beq[eq:eprob]
\bea{rcllll}
e_\joint(i)
&=&
\prob{\mbox{detect $i$ AND input $j\neq i$}}
&=&
\pout(i)-\pjoint(i,i)
\\&&&&\\
e_\cond(i)
&=&
\prob{\mbox{detect $j\neq i$ GIVEN input $i$}}
&=&
1-\poutin(i|i)
\\&&&&\\
e_\post(i)
&=&
\prob{\mbox{detect $i$ GIVEN input $j\neq i$}}
&=&
1-\pinout(i|i)
\eea
\eeq
Observe that each of these is the sum of the off-diagonal elements of
the corresponding probability matrices \refeq{probmats}. Being error
probabilities, they all range from zero to one:
\beq[eq:e range]
e_\joint(i),\
e_\cond(i),\
e_\post(i)\
\in
[0,1],\
i=1,\ldots,m
\eeq
As we will see shortly, it is convenient to express each of the error
probabilities in terms of $\poutin(i|j)$ and $p_i $. Using
\refeq{allprob} gives,
\beq[eq:eprob 1]
\bea{rcl}
e_\joint(i)
&=&
\sum_{j=1}^m\ \poutin(i|j)\ p_j  
-\poutin(i|i)\ p_i 
\\
e_\cond(i)
&=&
1-\poutin(i|i)
\\
e_\post(i)
&=&
\ds
1-\frac{\poutin(i|i)\ p_i }{\pout(i)}
=
1-\frac{\poutin(i|i)\ p_i }{\sum_{j=1}^m\ \poutin(i|j)\ p_j }
\eea
\eeq
The expression for $e_\post(i)$ is valid only if $\pout(i) \neq 0$
which is assumed.

%%%%%%%%%%%%%%%%%%%%%%%%%%%%%%%%%%
\subsection{Perfect and unambiguous detection}
\label{sec:perfumb}

{\em Perfect detection} occurs when the detector reads $i$ only if the
$i$th input is present. Thus the detector is correct all the time. In
this case the \apost probability matrix is identity which can only
occur when the conditional probability matrix is identity, \ie,
\beq[eq:perfect]
\poutin(i|j)
=
\del_{ij},\;
i,j=1,\ldots,m
\eeq
Under this condition, {\em all} the error probabilities in
\refeq{eprob 1} are simultaneously identically zero.  As might also be
expected, perfect performance is {\em independent} of the input
distribution $\seq{p_j }$. For quantum systems, this is possible if
and only if the input states are orthogonal \cite{P95}.

A weaker condition, referred to as {\em unambiguous detection}, occurs
when the detector either provides the correct answer or one that is
inconclusive with some probability (see, \eg,
\cite{E03p,E02}. This detector requires an additional
outcome corresponding to the inconclusive result. There are now $m+1$
detector outcomes, $\Dbf_{\rm out}=\set{i}{i=0,\ldots,m}$, where
outcome $0$ means the result is inconclusive. As before, for
$i=1,\ldots,m$, outcome $i$ means that input $i$ is declared to be
present. For the detector to be correct when $i,\ i=1,\ldots,m$ is
declared, the \apost probability of input $i$ given outcome $i$
must be 1. Equivalently, the submatrix of the conditional probability
matrix corresponding to the $m$ states is diagonal but not necessarily
identity, as in perfect detection. Thus \refeq{perfect} now becomes,
\beq[eq:imperfect]
\poutin(i|j)
=
\pb(i)
\del_{ij},\;
i,j=1,\ldots,m
\eeq
Under this condition, the \apost error probability is,
\beq[eq:unamb 1]
e_\post(i)
=
1-\frac{\poutin(i|i)\ p_i }{\sum_j\ \poutin(i|j)\ p_j }
=
1-\frac{\pb(i)\ p_i }{\pb(i)\ p_i }
=0
\eeq
for all $i=1,\ldots,m$, and the probability of an inconclusive result
is,
\beq[eq:pincl 0]
\pincl = 1 - \sum_{i=1}^m\ \pb(i) p_i 
\eeq
If the probability of an inconclusive result is non-zero, then this
detector is a type of {\em randomized detector}.  A detector designed
without this feature will be referred to as a {\em deterministic
detector}.  We will return to the problem of designing an unambiguous
and/or randomized detector in Section \refsec{umb}.

%%%%%%%%%%%%%%%%%%%%%%%%%%%%%%%%%
\subsection{Partial state detection}
\label{sec:partdetect}

It is often the case that not all the input states are to be detected.
We will show that it is not necessary to have a detector outcome for
all the states.  Consider the input set,
\beq[eq:inbig]
\Dbf_{\rm in}
=
\set{\rho_i,\ p_i}{i=1,\ldots,\ell}
\eeq
Suppose only the states $\rho_1,\ldots,\rho_k,\ k<\ell$ are to be
detected. The $\ell-k$ states that are not being detected can be
lumped into one state, the statistical mixture,
\beq
r = \sum_{i=k+1}^\ell\ p_i \rho_i
\eeq
with occurrence probability $\sum_{i=k+1}^\ell p_i$.  Thus the set
\refeq{inbig} of $\ell$ states can be replaced with the statistically
equivalent set of $k+1 < \ell$ states
\beq[eq:insmall]
\Dbf_{\rm in}
=
\seq{
(\rho_1,\ p_1),
\ldots,
(\rho_k,\ p_k),\
(r,\ \sum_{i=k+1}^\ell\ p_i)
}
\eeq
The detector then only requires $k+1$ outcomes, not $\ell$ outcomes.
To adhere to the previous notation, \eg, \refeq{omin}, define $m=k+1$.

An important application of the above procedure is detection of a
single pure state.  In this case the input state set, in the form of
\refeq{insmall}, becomes,
\beq[eq:omin pure]
\Dbf_{\rm in}
=
\left\{
(\psi \psi^*,\ 1-\bet),\; (r,\ \bet)
\right\}
\eeq
with the pure state $\psi\in\Cbf^n,\ \psi^*\psi=1$ occurring with
probability $1-\bet$ and the remaining states represented by the mixed
state $r\in\Cbfnn,\ r>0,\ \trace\ r=1$ occurring with probability
$\bet$. We use this example to illustrate the structure of the optimal
detector in some cases.

%%%%%%%%%%%%%%%%%%%%%%%%%%%%%%%%%%%
\subsection{Measures of performance}
\label{sec:metrics}

The goal is to design the detector to minimize the size of an error
probability. The size of the error is set by selecting a norm. Here we
will consider two common norms referred to as {\em average} and {\em
worst-case}. Since \refeq{e range} holds -- the errors are always
non-negative -- we can define the average error norm by
$\normav{e}=\sum_{i=1}^m\ w_i\ e(i)$ and the worst-case norm by
$\normwc{e}=\max_{i=1,\ldots,m}\ w_i\ e(i)$. These norms are {\em
weighted error probabilities}: the weights, $w_i \geq 0$, are selected
to emphasize specific outcomes -- a larger weight emphasizes the
desire to detect a particular state.  Table \ref{tab:e norm}
shows these norms for the specific error probabilities \refeq{eprob
1}.
%
%\beq[eq:e norm table]

\begin{table}[h]
\[
\renewcommand{\arraystretch}{3}
\bea{|c||c|c|}\hline
e & \normav{e} & \normwc{e}\\
\hline\hline
e_\joint
&
\ds\sum_i\
w_i
\left(
\pout(i)-\poutin(i|i)\ p_i 
\right)
&
\ds\max_i\
w_i
\left(
\pout(i)-\poutin(i|i)\ p_i 
\right)
\\
\hline
e_\cond
&
\ds\sum_i\
w_i
\left(
1-\poutin(i|i)
\right)
&
\ds\max_i\
w_i
\left(
1-\poutin(i|i)
\right)
\\
\hline
e_\post
&
\ds\sum_i\
\ds
w_i
\left(
1-\frac{\poutin(i|i)\ p_i }{\sum_j\ \poutin(i|j)\ p_j }
\right)
&
\ds\max_i\
w_i
\left(
1-\frac{\poutin(i|i)\ p_i }{\sum_j\ \poutin(i|j)\ p_j }
\right)
\\
\hline
\eea
\]
\caption{Norms of error probabilities.}
\label{tab:e norm}
\end{table}

%\eeq
%
In Table \ref{tab:e norm}, the performance measures 
$
\normav{e_\joint},\ 
\normav{e_\cond},\ 
\normwc{e_\joint},\ 
\normwc{e_\cond}
$
are convex functions of the elements of the conditional probability
matrix.  In the next section we will show that the conditional
probabilities are affine functions of the design parameters,
specifically, the elements of the POVM characterizing the
detector. Hence, these are convex functions of the design
parameters. Of these measures of performance, only
$\normav{e_\joint},\ \normav{e_\cond}$, or slight variations thereof,
have been addressed in the literature. Section \refsec{optav joint}
briefly describes the convex optimization problem associated with the
performance measure $\normav{e_\joint}$ and is to some extent a
partial review of known results, \eg, \cite{EMV02}.

The performance measure $\normwc{e_\post}$, which is the focus of this
paper, is a {\em quasiconvex} function of the conditional
probabilities, and hence, a quasiconvex function of the design
parameters, \eg, the POVM elements. As we will show in Section
\refsec{optwc post}, the optimal design can be obtained by
solving a convex optimization problem.

The performance measure $\normav{e_\post}$ is neither a convex nor
quasiconvex function, hence, only local solutions are guaranteed to be
found numerically.

%%%%%%%%%%%%%%%%%%%%%%%%%%%%%%
\iffalse
A detector which maximizes a measure of the conditional or joint
performance probability is referred to as a ML detector for Maximum
Likelihood.  A detector which maximizes a measure of the posterior
performance probability is referred to as a MAP detector for Maximum A
Posteriori.
\fi
%%%%%%%%%%%%%%%%%%%%%%%

%%%%%%%%%%%%%%%%%%%%%%%%%%%%%%%%%%%%%%%%%%%%%%%%%%%%%
\section{Detector as a POVM}
\label{sec:povm}

We start with the assumption that the detector can be completely
described by a POVM (positive operator valued measure) with matrix
elements $\set{O_i\in\Cbfnn}{i=1,\ldots,m}$ which, by definition,
satisfy,\footnote{
The notation $X \geq 0$ or $X>0$ means that $X=X^*$ and all the
eigenvalues of $X$ are, respectively, non-negative or strictly
positive.
}
\beq[eq:povm]
\sum_{i=1}^m\\ O_i = I_n,
\;\;\;\;
O_i \geq 0,\ i=1,\ldots,m
\eeq
In consequence, designing an {\em optimal} detector means selecting
the matrices that form the POVM to minimize a selected performance
measure from Table \ref{tab:e norm}. To this end we express the error
probabilities in terms of the problem data $\seq{\rho_i,\ p_i }$ and
the design variables $\seq{O_i}$. First, the conditional probability
of detecting $i$ given state $j$ is,
\beq[eq:cond 1]
\poutin(i|j) = \trace\ O_i \rho_j
\eeq
From \refeq{allprob}, the total probability of detector event $i$ is then,
\beq[eq:palf1]
\pout(i) 
=
\sum_{j=1}^m\ \poutin(i|j)\ p_j 
=
\sum_{j=1}^m\  (\trace\ O_i \rho_j) p_j 
=
\trace\ O_i\ \rho
\eeq
where $\rho$ is the statistical mixture of all the possible input states,
\beq[eq:rho]
\rho = \sum_{j=1}^m \ p_j \ \rho_j
\eeq
Throughout we make the assumption that,
\beq[eq:rhopos]
\rho > 0
\eeq
This is a not a limiting condition; it is easily satisfied in most
practical situations and if necessary can be overcome by restricting
attention to the range space of $\rho$.

The error probabilities in \refeq{eprob 1} can now be expressed as
follows for $i=1,\ldots,m$
\beq[eq:eprob 2]
\bea{rcl}
e_\joint(i)
&=&
\ds
\trace\ O_i \left( \rho - p_i  \rho_i \right)
\\
e_\cond(i)
&=&
\ds
1-\trace\ O_i \rho_i
\\
e_\post(i)
&=&
\ds
1-\frac{p_i \ \trace\ O_i \rho_i}{\trace\ O_i \rho} 
\eea
\eeq
Observe that $e_\post(i)$ is meaningful only if $\trace\ O_i \rho
>0$. Since $O_i \geq 0$ from \refeq{povm}, it follows that if the
mixed state $\rho>0$, then $\trace\ O_i \rho =0$ only when $O_i=0$
which is a pathological case.
%The issue is addressed in more detail in the appendix. % \refsec{cvxopt}.

Using \refeq{eprob 2}, the entries in Table \ref{tab:e norm} 
are given explicitly as shown in Table \ref{tab:e norm povm}. 
The first observation to make is that the POVM matrices $\seq{O_i}$
form a convex set \refeq{povm}. As already stated, since $e_\joint(i)$
and $e_\cond(i)$ are affine functions of $O_i$, it follows that these
errors are both convex functions of the $O_i$ matrices. Further, since
all norms are convex functions, the performance measures
$\normav{e_\joint},\ \normwc{e_\joint}$ and $ \normav{e_\cond},\
\normwc{e_\cond}$ are all convex functions of the POVM
matrices. Again, we note that $\normwc{e_\post}$ is a quasiconvex
function of $\seq{O_i}$ and $\normav{e_\post}$ is not convex.
Therefore, minimizing any of these (quaisi)convex measures over the
POVM matrices can be cast as a convex optimization problem.

%
%\beq[eq:e norm table povm]

\begin{table}[h]
\[
\renewcommand{\arraystretch}{3}
\bea{|c||c|c|}\hline
e & \normav{e} & \normwc{e}\\
\hline\hline
e_\joint
&
\ds\sum_i\
w_i
\left(
\trace\ O_i \left( \rho - p_i  \rho_i \right)
\right)
&
\ds\max_i\
w_i
\trace\ O_i \left( \rho - p_i  \rho_i \right)
\\
\hline
e_\cond
&
\ds\sum_i\
w_i
\left(
1-\trace\ O_i \rho_i
\right)
&
\ds\max_i\
w_i
\left(
1-\trace\ O_i \rho_i
\right)
\\
\hline
e_\post
&
\ds\sum_i\
w_i
\left(
1-\frac{p_i \ \trace\ O_i \rho_i}{\trace\ O_i \rho} 
\right)
&
\ds\max_i\
w_i
\left(
1-\frac{p_i \ \trace\ O_i \rho_i}{\trace\ O_i \rho} 
\right)
\\
\hline
\eea
\]
\caption{Norms of error probabilities as functions of POVM elements.}
\label{tab:e norm povm}
\end{table}

%\eeq
%

\subsubsection*{Optimality conditions}

Lagrange Duality Theory \cite[Ch.5]{BoydV:04} provides a means for
establishing a lower bound on the optimal objective value,
establishing conditions of optimality, and providing, in some cases, a
more efficient means to numerically solve the original problem.  In
the sections to follow we will present the optimality conditions in a
form which involves only the problem data, $\rho_i,\ p_i,\
i=1,\ldots,m$, and the design variables, the POVM matrices, $O_i,\
i=1,\ldots,m$. The details are presented in the Appendix.  The
optimality conditions can also be used for determining if a known POVM
set is optimal, what the authors in \cite{AudD:03} call: ``testing an
Ansatz.'' Such a POVM could be obtained from some analytic means, from
data, or from imagination.

%%%%%%%%%%%%%%%%%%%%%%%%%%%%%%%%%%%%%%%
\subsubsection*{Implementation of a POVM}
%\label{sec:imp povm}

As shown in \cite[\S 2.2.8]{NielsenC:00}, any POVM can be implemented
by a unitary matrix in an expanded space together with rank-one
projective measurements in the natural basis on the ancilla
outputs. Some general implementations of a POVM are also presented in
\cite{LloydV:00}. Realizing the resulting unitary and rank-one
projections with {\em specific} physical components is, in general, a
more difficult problem.

%%%%%%%%%%%%%%%%%%%%%%%%%%%%%%%%%%%%%%%%%%
\section{Optimal average joint performance}
\label{sec:optav joint}

In this section we briefly discuss optimal detector design for
$\normav{e_\joint}$. This problem, with slight variations, has been
essentially completely analyzed in \cite{EMV02}. The presentation
here is primarily to illustrate a few of the ideas which repeatedly
occur.  Following this, the main focus of this paper, presented in
Section \refsec{optwc post}, is on detector design for
$\normwc{e_\post}$, the optimal worst-case \apost design.

A detector which minimizes the objective $\normav{e_\joint}$ in Table
\ref{tab:e norm povm} is obtained by solving the following
optimization problem for the POVM matrices $\seq{O_i}$:
\beq[eq:optjoint av]
\bea{ll}
\mbox{minimize}
&
\normav{e_\joint}
=
\sum_{i=1}^m\
w_i
\Big(
\trace\ O_i\ (\rho - p_i  \rho_i)
\Big)
\\
\mbox{subject to}
&
\sum_{i=1}^m\ O_i = I_n,
\;\;\;
O_i \geq 0,\ i=1,\ldots,m
\eea
\eeq
This problem was addressed in \cite{EMV02} for equal weights,
$w_i=1$, where the objective becomes $1-\sum_i\ p_i \trace\ O_i
\rho_i$. As observed in \cite{EMV02}, problem \refeq{optjoint
av}, with or without equal weights, is a {\em semidefinite program}
(SDP) \cite[\S 4.6.2]{BoydV:04}. An SDP is a generalization of linear
programming where the linear inequalities are replaced with matrix
inequalities.  Although it does not make any physical sense, if the
$\seq{O_i}$ are constrained to be diagonal, then the problem reduces
to a {\em linear programming problem}.

\subsubsection*{Optimality conditions}

As derived in Appendix \refsec{oc joint}, any feasible POVM, \ie, any
set of $O_i\in\Cbfnn,\ i=1,\ldots,m$ which satisfy \refeq{povm}, is
optimal for problem \refeq{optjoint av} if and only if,
\beq[eq:optavjoint cond]
\bea{rcl}
A_i - \sum_{j=1}^m\ A_j O_j &\geq& 0,\;
i=1,\ldots,m
\\
\left( A_i - \sum_{j=1}^m\ A_j O_j \right) O_i &=& 0,\;
i=1,\ldots,m
\eea
\eeq
with all the problem data in the matrices,
\beq[eq:optavjoint amat]
A_i = w_i \left( \rho - p_i \rho_i \right),\;
i=1,\ldots,m
\eeq

\subsubsection*{Two state detection}

As an application consider the two state detection problem using the
state set \refeq{omin pure}. For equal weights $w_1=w_2=1$, the data
matrices are,
\beq[eq:a1a2]
\bea{rcl}
A_1 &=& \rho-(1-\bet)\psi\psi^* = \bet r 
\\
A_2 &=& \rho-\bet r = (1-\bet)\psi\psi^*
\eea
\eeq
Using $O_1+O_2=I$, the optimality conditions become,
\beq[eq:oc a1a2]
\bea{ll}
A O_2 \geq 0 & A O_2 O_1 =0
\\
A O_1 \leq 0 & A O_1 O_2 =0
\eea
\eeq
with
\beq[eq:oc a]
A = A_1 - A_2 = \bet r - (1-\bet)\psi\psi^*
\eeq
Since $A$ is Hermitian it can be decomposed as,
\beq[eq:uomu]
A 
=
\left[U_+\ U_- \right]\
\left[
\bea{cc} \Om_+ & 0
\\ 
0 & \Om_- \eea
\right]\
\left[
\bea{c} U_+^* \\ U_-^* \eea
\right]
\eeq
where $U=[U_+\ U_-]\in\Cbfnn$ is unitary and $(\Om_+\geq 0,\ \Om_-\leq
0)$ are diagonal matrices consisting, respectively, of the positive
and negative eigenvalues of $A$. Make the choice,
\beq[eq:o1o2]
O_1 = U_- U_-^*,
\;\;
O_2 = U_+ U_+^*
\eeq
This is a feasible POVM set because $U$ is unitary. This choice also
satisfies the optimality conditions \refeq{oc a1a2}. Specifically,
$AO_1=U_-\Om_-U_-^* \leq 0$, $AO_2=U_+\Om_+U_+^* \geq 0$, and because
unitary $U$ requires $U_-^*U_+=0$, it follows that $O_1
O_2=U_-U_-^*U_+U_+^*=0$.  After some algebra, the optimal objective
value is found to be,
\beq[eq:optobj]
\normav{e_\joint^\opt}
=
\trace(O_1A_1+O_2A_2)
=
\bet-\trace\ \Om_+
\eeq
As a further illustration, assume that $r$ is completely randomized,
that is, $r=I_n/n$. In this case $U$ can be chosen such that the pure
state in refeq{omin pure} has the decomposition $\psi\psi^*=U\
\diag(0,\ldots,0,1)\ U^*$. It then follows that:
\beq[eq:ill]
\bea{rcl}
A &=& \diag\left( \bet/n,\ldots,\bet/n,-1+\bet(1+1/n) \right)
\\
\rho &=& \diag\left( \bet/n,\ldots,\bet/n,1-\bet(1-1/n) \right)
\\
\Om_+ &=& \diag\left( \bet/n,\ldots,\bet/n \right)
\\
\Om_- &=& -1+\bet(1+1/n)
\eea
\eeq
Observe that $\Om_- \leq 0$ if and only if $\bet \leq n/(1+n)$.  In
general, as we show below, the assumption that $A$ has both positive
and negative eigenvalues places a limit on the size of $\bet$.

Since $\Om_+$ is $n-1 \times n-1$, we get $\trace\ \Om_+=\bet (n-1)/n$,
and hence the objective value becomes
\beq[eq:optobj 1]
\normav{e_\joint^\opt}
=
\bet/n
\eeq
For this detector the corresponding \apost probabilities are
\beq[eq:apost]
\bea{rcl}
\pinout(1|1)
&=&
\ds
\frac{(1-\bet)\trace\ O_1 \psi\psi^*}{\trace\ O_1\rho}
=
\frac{1-\bet}{1-\bet(1-1/n)}
\\
\pinout(2|2)
&=&
\ds
\frac{\bet\trace\ O_2 r}{\trace\ O_2\rho}
=
\frac{(n-1)\bet}{n-\bet}
\eea
\eeq
This result shows that as $n\to\infty$, $\normav{e_\joint}\to 0$,
$\pinout(1|1)\to 1$, and $\pinout(2|2)\to\bet$. Thus, if the state
dimension is large and the statistical mixture of the residual states
tends to average out to a random distribution over all states, then
the probability of detecting a single pure state is very high.

\paragraph{Restrictions on $\bet$}

From \refeq{oc a}, $A$ has non-negative eigenvalues ($A\geq 0$) only
if $\bet r \geq (1-\bet)\psi\psi^*$, or equivalently, if,
\beq[eq:bet0]
\bet \geq \bet_0 
= \frac{\psi^*r^{-1}\psi}{1+\psi^*r^{-1}\psi}
\eeq
Hence, for $\bet < \bet_0$, $A$ will have both positive and negative
eigenvalues. If as in the above example, $r=I_n/n$, then
$\psi^*r^{-1}\psi=n$ and thus $\bet_0=n/(1+n)$.

Suppose that $A>0$ ($\bet > \bet_0$). Then the only way to satisfy the
optimality conditions \refeq{oc a1a2} is to set
\beq[eq:o1o2 bet0]
O_1=0,
\;\;
O_2=I_n
\eeq
The optimal objective value is now
\beq[eq:optobj bet0]
\normav{e_\joint^\opt}
=
\trace(A_2)
=
1-\bet
\eeq
This is just the occurrence probability of the pure state $\psi$;
essentially the detector does nothing.  Observe that similar remarks
can be made when $A<0$.

%%%%%%%%%%%%%%%%%%%%%%%%%%%%%%%%%
\section{Optimal worst-case \apost design}
\label{sec:optwc post}

In this section the detector is designed to minimize the objective
$\normwc{e_\post}$ in Table \ref{tab:e norm povm}. This requires
solving the following optimization problem for the POVM matrices
$\seq{O_i}$:
\beq[eq:optpost]
\bea{ll}
\mbox{minimize}
&
\ds
\normwc{e_\post}
=
\max_{i=1,\ldots,m}\
w_i
\left(
1-
\frac{
p_i \ \trace\ O_i\ \rho_i
}{
\trace\ O_i \rho
}
\right)
\\
\mbox{subject to}
&
\sum_{i=1}^m\ O_i = I_n,
\;\;\;
O_i \geq 0,\;
i=1,\ldots,m
\\
&
\trace\ O_i \rho >0,\;
i=1,\ldots,m
\eea
\eeq
As shown in \cite[\S 4.3.2]{BoydV:04}, the objective function,
$\normwc{e_\post}$, is a maximum over a set of {\em quasiconvex
functions} each with domain $\trace\ O_i\rho>0,\ \forall i$, and hence, is a
quasiconvex function over the domain $\set{\trace\
O_i\rho>0}{i=1,\ldots,m}$.  Since the POVM matrices $\seq{O_i}$ form a
convex set, \refeq{optpost} is a {\em quasiconvex optimization
problem} in the POVM matrices. Technically this means that for any
positive scalar $\del$, the {\em sublevel sets} of POVMs 
\beq[eq:sublevel]
\set{O_i,\ \trace\ O_i\rho>0,\ \forall i}
{\normwc{e_\post} \leq \del}
\eeq
are convex. To see that these sets are convex in this case, observe
that for POVMs in the domain $\trace\ O_i\rho>0,\ \forall i$, the
sublevel sets are equivalently,
\beq[eq:sublevel 1]
\trace\ O_i A_i(\del) \leq 0
\eeq
with 
\beq[eq:optwcpost amat]
A_i(\del) = (w_i-\del) \rho - w_i p_i \rho_i,\;
i=1,\ldots,m
\eeq
The sets defined by \refeq{sublevel 1} are affine in the POVM
elements, and hence, are convex sets.  We will refer to the matrices
$A_i(\del)$ as the {\em data matrices}.

This effectively shows (see also Appendix \refsec{oc apost}) that
\refeq{optpost} is equivalent to,
\beq[eq:optpost a]
\bea{ll}
\mbox{minimize}
&
\del
\\
\mbox{subject to}
&
\sum_{i=1}^m\ O_i = I_n,
\;\;\;
O_i \geq 0,\;
i=1,\ldots,m
\\
&
\trace\ O_i A_i(\del) \leq 0,\;
i=1,\ldots,m
\eea
\eeq
The optimization variables for \refeq{optpost a} are now the real
positive scalar $\del$ as well as the POVM matrices $\seq{O_i}$.
Observe that \refeq{optpost a} does not include the constraint set
$\trace\ O_i \rho >0,\ i=1,\ldots,m$. Since $\rho>0$, the only way
this constraint can be violated is if a POVM element is zero. If this
occurs then the problem is ill-posed and most likely that POVM element
can be eliminated. Hence, from now on we do not explicitly state this
constraint.

As shown in \cite[\S 4.2.5]{BoydV:04} and described in Appendix
\refsec{oc apost}, a solution to the quasiconvex optimization problem
\refeq{optpost} or \refeq{optpost a} can be obtained by solving a
series of convex feasibility problems together with a bisection
method.

%%%%%%%%%%%%%%%%%%%%%%%%%%%%%%%%%%%%%%%%%%%
\subsubsection*{Optimality conditions}

As derived in Appendix \refsec{oc apost}, any feasible POVM
\refeq{povm} is optimal if and only if there exist real constants
$\del^\opt$ and $\lam_i,\ i=1,\ldots,m$ such that,
\beq[eq:optwcpost cond]
\bea{rcl}
\lam_i &>& 0,\; i\in S
\\
\lam_i &=& 0,\; i\not\in S
\\
\sum_{i=1}^m\ \lam_i &=& 1
\\
\lam_i A_i(\del^\opt)-\sum_{j=1}^m\ \lam_j A_j(\del^\opt) O_j &\geq& 0,\;
i=1,\ldots,m
\\
\left( 
\lam_i A_i(\del^\opt)-
\sum_{j=1}^m\ \lam_j A_j(\del^\opt)O_j
\right)
O_i &=& 0,\;
i=1,\ldots,m
\eea
\eeq
The index set $S$ consists only of those indices where the optimal
$\del^\opt$ in \refeq{optpost a} is achieved. Thus $S$ is equivalently
expressed by,
\beq[eq:idx post]
\bea{rcl}
S &=& \set{i=1,\ldots,m}{\trace\ O_i A_i(\del^\opt) =0}
\\&&\\
&=&
\ds
\set{i=1,\ldots,m}
{
\del^\opt=w_i
\left(
1-
\frac{
p_i \ \trace\ O_i\ \rho_i
}{
\trace\ O_i \rho
}
\right)
}
\eea
\eeq
Some special cases follow.

\subsubsection*{Equal weights: one active linear constraint}

For equal weights, $w_i=1,\ \forall i$, \refeq{optpost} can be
expressed equivalently by,
\beq[eq:optpost wtpart]
\bea{ll}
\mbox{maximize}
&
\gam
\\
\mbox{subject to}
&
\ds
\pinout(i|i)
=
\frac{p_i \ \trace\ O_i\ \rho_i}{\trace\ O_i \rho}
\geq \gam,\;
i=1,\ldots,m
\\
&
\sum_{i=1}^m\ O_i = I_n,
\;\;\;
O_i \geq 0,\;
i=1,\ldots,m
\eea
\eeq
Clearly $\gam=1-\del$ with $\del$ from \refeq{optpost a}.  Let
$\gam^\opt$ denote the optimal objective value in \refeq{optpost
wtpart}. Suppose only one linear constraint is active, that is, for
$i=k$, $\trace\ O_k A_k(\gam^\opt) = 0$ and for $i\neq k$, $\trace\
O_i A_i(\gam^\opt) < 0$. Then, as shown in Appendix \refsec{oc apost},
the optimality conditions \refeq{optwcpost cond} become,
\beq[eq:optwcpost cond 2]
\bea{rcl}
A_k(\gam^\opt) \left(I-O_k\right) &\geq& 0
\\
A_k(\gam^\opt) O_k &\leq& 0
\\
A_k(\gam^\opt) \left(I-O_k\right) O_k &=& 0
\\
A_k(\gam^\opt) O_k O_i &=& 0,\;
i\neq k
\\
\trace\ O_k A_k(\gam^\opt) &=& 0
\eea
\eeq
with $\gam^\opt$ given by,
\beq[eq:optval]
\gam^\opt =
p_k \ \sigmax(\rho^{-1/2} \rho_k \rho^{-1/2}) 
\eeq
where $\sigmax(\cdot)$ is the maximum singular value of the matrix
argument.  (Note that $\rho^{-1/2}$ exists because $\rho>0$ is assumed
\refeq{rhopos}). Since only one constraint is assumed active,
\refeq{optval} is equivalent to,
\beq[eq:optval 1]
\gam^\opt
=
\min_{i=1,\ldots,m}\ 
p_i \ \sigmax(\rho^{-1/2} \rho_i \rho^{-1/2}) 
\eeq
If the input states are {\em pure}, that is,
\beq[eq:rhoi pure]
\rho_i = \psi_i \psi_i^*,\;
i=1,\ldots,m
\eeq
with $\psi\in\Cbf^n,\ \psi^*\psi=1$, then \refeq{optval} becomes,
\beq[eq:optval pure]
\gam^\opt
=
\min_{i=1,\ldots,m}\ 
p_i \ \psi_{i}^* \rho^{-1} \psi_{i}
\eeq
%

%%%%%%%%%%%%%%%%%%%%%%%%%%%%%%%%%%%%%%%%%%%%%%%%%%%%%
\subsubsection*{Single pure state detection}

Consider again the input set \refeq{omin pure} where the goal is to
detect $\psi$. With the weights set to $w_1=1,\ w_2=0$, the data
matrices are,
\beq[eq:a1gam]
\bea{rcl}
A_1(\gam) &=& \gam\rho-(1-\bet)\psi\psi^*
\\
A_2(\del) &=& -\del \rho
\eea
\eeq
with $\gam=1-\del$. Unless $\gam^\opt=1\ (\del^\opt=0)$, it follws
that $\trace\ O_2A_2(\del^\opt)=-\del^\opt\trace\ O_2\rho<0$. Thus,
the only active constraint is $\trace\ O_1A_1(\gam^\opt)=0$ which
makes the index set $S$ the singleton $S=\seq{1}$, and hence,
\refeq{optwcpost cond 2}-\refeq{optval} applies.  Using the Matrix
Inversion Lemma to compute $\psi^*\rho^{-1}\psi$ with
$\rho=(1-\bet)\psi\psi^*+\bet r$ gives,
\beq[eq:gamopt]
\gam^\opt
=
(1-\bet) \psi^* \rho^{-1} \psi
= 
\frac{1-\bet}{1-\bet(1-1/\psi^* r^{-1} \psi)}
\eeq
Observe that $\gam^\opt$ increases as $\psi^* r^{-1} \psi$ increases.
If $\psi$ is close to a singular vector of $r$ which has a very small
singular value, then $\psi^* r^{-1} \psi$ will be large, and hence,
$\gam^\opt\approx 1$. This can be construed as an approximate
orthogonality condition.  In the special case when $r=I_n/n$, then
$\psi^* r^{-1} \psi=n$, and hence,
\beq[eq:gamopt eye]
\gam^\opt
=
\frac{1-\bet}{1-\bet(1-1/n)}
\eeq
This is exactly the result in \refeq{apost}, which in general is not
to be expected.

The (two) POVM elements $O_1$ and $O_2$ associated with the two state
input set \refeq{omin pure} can be directly calculated from the
optimality conditions \refeq{optwcpost cond 2}.  Using the fact that
$O_1+O_2=I$, the optimality conditions \refeq{optwcpost cond 2} become:
\beq[eq:ocwc]
\bea{ll}
A_1(\gam^\opt)O_2 \geq 0
&
A_1(\gam^\opt)O_2 O_1 = 0
\\
A_1(\gam^\opt)O_1 \leq 0
&
A_1(\gam^\opt)O_1 O_2 = 0
\eea
\eeq
Observe that because $\lam_2=0$, the data matrix $A_2(\del^\opt)$
plays no part in the optimality conditions.  Using $\gam^\opt$
from \refeq{gamopt} makes $\rank\ A_1(\gam^\opt)=n-1$, and hence has
the decomposition,
\beq[eq:svd aopt]
\bea{l}
A_1(\gam^\opt)
=
\gam^\opt \rho-(1-\bet)\psi\psi^*
=
\left[U_+\ U_0 \right]\
\left[
\bea{cc} \Om_+ & 0_{n-1} 
\\ 
0_{n-1}^T & 0 \eea
\right]\
\left[
\bea{c} U_+^* \\ U_0^* \eea
\right]
\\
\\
\Om_+ = {\bf diag}(\om_1,\ldots,\om_{n-1}),
\;
\om_1 \geq \om_2 \geq \cdots \geq \om_{n-1} \geq 0,
\;\; \om_1>0
\eea
\eeq
for unitary $[U_+\ U_0]\in\Cbfnn$ with $U_+\in\Cbf^{n\times n-1}$ and
$U_0\in\Cbf^{n\times 1}$. Setting,
\beq[eq:povm pure]
O_1 = U_0 U_0^*,
\;\;
O_2 = U_+ U_+^*
\eeq
gives $A_1(\gam^\opt)O_2=U_+\Om_+U_+^* \geq 0,\ A_1(\gam^\opt)O_1=0,\
O_1O_2=0$, thus satisfying the optimality conditions \refeq{ocwc}.
Observe also that $O_1$ is a rank $1$ projector, and $O_2$ is a rank
$n-1$ projector.

If $r=I_n/n$, then the \apost probabilities are exactly the same as
given by \refeq{apost}; again, this is not the case in general.

%%%%%%%%%%%%%%%%%%%%%%%%%%%%%%%%%
\subsubsection*{Single state detection with pure residual state}

In the previous example, as long as the residual state $r>0$, then it
it is not possible to make $\gam^\opt=1$. To see this, observe that
$A_1(\gam^\opt=1)= \rho-(1-\bet)\psi\psi^*=\bet r > 0$, and hence has
only positive eigenvalues. Thus, the optimality conditions can only be
satisfied with $O_1=0,\ O_2=I$. (Effectively $U_0$ in \refeq{svd aopt}
is null.) This choice of $\gam^\opt$ is therefore infeasible.

Now consider the input set,
\beq[eq:omin rpure]
\Dbf_{\rm in}
=
\left\{
(\rho_0,\ 1-\bet),\; (\phi\phi^*,\ \bet)
\right\}
\eeq
with the pure residual state $\phi\in\Cbf^n,\ \phi^*\phi=1$ occurring
with probability $\bet$ and the state to be detected
$\rho_0\in\Cbfnn,\ \rho \geq 0$ occurring with probability $1-\bet$.
In this case for $\gam^\opt=1$, we get,
\beq[eq:a1 rpure eig]
\bea{rcl}
A_1(\gam^\opt=1)
&=& 
\rho-(1-\bet)\rho_0
\\
&=&
\bet \phi\phi^*
=
\bet\
\left[U_+\ U_0 \right]\
\left[
\bea{cc} 
1 & 0_{n-1}^T 
\\ 
0_{n-1} & 0_{n-1\times n-1} 
\eea
\right]\
\left[
\bea{c} U_+^* \\ U_0^* \eea
\right]
\eea
\eeq 
with $U_+\in\Cbf^{n\times 1},\ U_0\in\Cbf^{n\times n-1}$. The choice
$O_1=U_0 U_0^*,\ O_2=U_+U_+^*$ satisfies the optimality conditions.
Hence, perfect deterministic detection of a single state, pure or
mixed, is possible if the residual state is pure.

%%%%%%%%%%%%%%%%%%%%%%%%%%%%%%%%%%%%%%%%%%%
\subsection{Noisy measurements}
\label{sec:noisy
 meas}

The optimal detector design problem can be modified to handle a
``noisy'' set of measurements.  In general there can be more noisy
measurements than noise-free measurements.  Consider, for example, a
photon detection device with two photon-counting detectors.  If both
are noise-free, meaning, perfect efficiency and no dark count probability,
then, provided one photon is always present at the input of the
device, there only two possible outcomes: $\seq{10,\ 01}$. If,
however, each detector is noisy, then either or both detectors can
misfire or fire even with a photon always present at the input. Thus in
the noisy case there are four possible outcomes: $\seq{10,\ 01,\ 11,\
00}$.

As before, let $\seq{O_i}$ denote the $m$ noise-free POVM
matrices. Now let $\seq{O^\noisy_i}$ denote the $\mh$ noisy measurements
with $\mh \geq m$. The noisy measurements can be expressed as,
\beq[eq:oalfmn]
O^\noisy_i = \sum_{j=1}^m \ 
\nu_{ij}\
O_{j},\
i=1,\ldots,\mh
\eeq
The $\seq{ \nu_{i} }$ represents the noise in the measurement,
specifically, the conditional probability that $i$ is measured given
the noise-free outcome $j$. Since $\sum_{i=1}^m\ \nu_{ij}=1,\
\forall j$, it follows that the noisy set $\seq{O^\noisy_i}$ is also
a POVM. Thus,
\beq[eq:povmn]
\sum_{i=1}^\mh\\ O^\noisy_i = I_n,
\;\;\;\;
O^\noisy_i \geq 0,\ i=1,\ldots,\mh
\eeq
In matrix form,
\beq[eq:povm2]
\left[
\bea{c}
O^\noisy_1 \\ \vdots \\ O^\noisy_\mh
\eea
\right]
=
\left[
\bea{ccc}
\nu_{11}\ I_n & \cdots & \nu_{1m}\ I_n 
\\
\vdots & \vdots & \vdots
\\
\nu_{\mh 1}\ I_n & \cdots & \nu_{\mh m}\ I_n 
\eea
\right]\
\left[
\bea{c}
O_1 \\ \vdots \\ O_m
\eea
\right]
\eeq
When the equivalent noisy POVM matrices, $\{ O^\noisy_i \}$, are
inserted into \refeq{optpost}, either objective function retains the
same form with the $\{ O^\noisy_i \}$ replacing the $\{ O_i \}$.  The
design variables are still the noise-free POVM matrices
$\seq{O_i}$. Since the noisy POVM matrices, $\{ O^\noisy_i \}$, are
linear in the noise-free POVM matrices, $\{ O_i \}$, the design
problems in Table \ref{tab:e norm povm} remain convex or quasiconvex
optimization problems over the noise-free POVM matrices
$\seq{O_i}$. 

%%%%%%%%%%%%%%%%%%%%%%%%%%%%
\subsubsection*{Optimal worst-case \apost performance with
noisy measurements}

With noisy measurements, \refeq{optpost} becomes,
\beq[eq:optpost 1]
\bea{ll}
\mbox{minimize}
&
\normwc{e_\post}
=
\ds
\max_{i=1,\ldots,m}\
w_i
\left(
1-\frac{
p_i \ \trace\ O^\noisy_i\ \rho_i
}{
\trace\ O^\noisy_i \rho
}
\right)
\\
\mbox{subject to}
&
O^\noisy_i 
= 
\sum_{j=1}^m \ 
\nu_{ij}\
O_{j},
\;\;
i=1,\ldots,\mh
\\
&
\sum_{i=1}^m\ O_i = I_n,
\;\;\;
O_i \geq 0,\ i=1,\ldots,m
\eea
\eeq
Under the assumption that $\trace\ O_i^\noisy \rho>0,\forall i$,
\refeq{optpost 1} is equivalent to,
\beq[eq:optpostnoise a]
\bea{ll}
\mbox{minimize}
&
\del
\\
\mbox{subject to}
&
O^\noisy_i 
= 
\sum_{j=1}^m \ 
\nu_{ij}\
O_{j},
\;\;
i=1,\ldots,\mh
\\
&
\sum_{i=1}^m\ O_i = I_n,
\;\;\;
O_i \geq 0,\;
i=1,\ldots,m
\\
&
\trace\ O_i^\noisy A_i(\del) \leq \del,\;
i=1,\ldots,m
\eea
\eeq
The optimization variables for \refeq{optpost 1} are now the real
positive scalar $\del$ as well as the noise-free POVM matrices
$\seq{O_i}$.  The data matrices, $A_i(\del)$, are given by
\refeq{optwcpost amat}.

%%%%%%%%%%%%%%%%%%%%%%%%%%%%%%%%%%%%%%%
\subsubsection*{Optimality conditions}

As derived in the Appendix \refsec{oc apost noise}, any feasible POVM
\refeq{povm} is optimal if and only if there exist real constants
$\lam_i,\ i=1,\ldots,m$ such that,
\beq[eq:optwcpostnoise cond]
\bea{rcl}
\lam_i &>& 0,\; i\in S
\\
\lam_i &=& 0,\; i\not\in S
\\
\sum_{i=1}^m\ \lam_i &=& 1
\\
\lam_i A_i(\del^\opt,\nu)-
\sum_{j=1}^m\ \lam_j A_j(\del^\opt,\nu) O_j &\geq& 0,\;
i=1,\ldots,m
\\
\left( 
\lam_i A_i(\del^\opt,\nu)-
\sum_{j=1}^m\ \lam_j A_j(\del^\opt,\nu)O_j
\right)
O_i &=& 0,\;
i=1,\ldots,m
\\
\sum_{i=1}^m\ \trace\ A_i(\del^\opt,\nu) O_i &=& 0
\eea
\eeq
with
\beq[eq:optwcpostnoise amat]
A_i(\del,\nu) = \sum_{j=1}^m\ \lam_j \nu_{ji} A_j(\del),\;
i=1,\ldots,m
\eeq
and the index set $S$ given by,
\beq[eq:idx post noisy]
S = \set{i=1,\ldots,m}{\trace\ O_i A_i(\del^\opt,\nu) =0}
\eeq
where 
$\del^\opt$ is the optimal objective value from \refeq{optpostnoise a}.

%%%%%%%%%%%%%%%%%%%%%%%%%%%%%%%%%%%%%%%%%%%%%%%%%%%%%
\subsubsection*{Single pure state detection}

Consider again the input set \refeq{omin pure} with weights $w_1=1,\
w_2=0$. Suppose the measurement noise matrix is,
\beq[eq:numat]
\nu
=
\left[
\bea{cc} 
1-\nu_0 & \nu_0
\\ 
\nu_0 & 1-\nu_0
\eea 
\right]
\eeq
Assuming the only active constraint to be $\trace\ O_1^\noisy
A_1(\del)=0$, then the multipliers are $\lam_1=1,\ \lam_2=0$ and the
optimality conditions become:
\beq[eq:ocwc noise]
\bea{ll}
(1-2\nu_0) A_1(\gam^\opt_\noisy)O_2 \geq 0
&
A_1(\gam^\opt_\noisy)O_2 O_1 = 0
\\
(1-2\nu_0) A_1(\gam^\opt_\noisy)O_1 \leq 0
&
A_1(\gam^\opt_\noisy)O_1 O_2 = 0
\\
\ds
\trace\ A_1(\gam^\opt_\noisy)
\left( O_1+\frac{\nu_0}{1-\nu_0} O_2 \right)
=0
&
\eea
\eeq
with $A_1(\gam)$ from \refeq{a1gam}. Assume further that
$\nu_0<1/2$. Then the matrix inequalities in \refeq{ocwc noise} are
the same as in \refeq{ocwc}, namely, 
\beq[oc matineq]
A_1(\gam^\opt_\noisy)O_2 \geq 0,
\;\;
A_1(\gam^\opt_\noisy)O_1 \leq 0
\eeq
Now again introduce the decomposition,
\beq[eq:uomu noise]
A_1(\gam^\opt_\noisy) 
=
\gam^\opt_\noisy \rho - (1-\bet)\psi\psi^*
=
\left[U_+\ U_- \right]\
\left[
\bea{cc} \Om_+ & 0
\\ 
0 & \Om_- \eea
\right]\
\left[
\bea{c} U_+^* \\ U_-^* \eea
\right]
\eeq
where $(\Om_+,\ \Om_-)$ are diagonal matrices consisting,
respectively, of the positive and negative eigenvalues of
$A_1(\gam^\opt_\noisy)$.  As in the previous examples, make the
choice,
$
O_1 = U_- U_-^*,
\;\;
O_2 = U_+ U_+^*
$
The matrix inequalities and equalities in the optimality conditions
are satisfied by this choice. The scalar (trace) condition is
satisfied provided that,
\beq[eq:ocwc noise 1]
\trace\ \Om_- = - \frac{\nu_0}{1-\nu_0} \trace\ \Om_+
\eeq
The noise-free case, $\nu_0=0$, requires that $\trace\ \Om_-=0$, which
means that $\Om_-=0$. This is the condition for the decomposition in
\refeq{svd aopt} which can only occur for $\gam^\opt_\noisy=\gam^\opt$ from
\refeq{gamopt}. When noise is present, $\nu_0>0$, it is necessary that
$\Om_-<0$, and hence, $\gam^\opt_\noisy < \gam^\opt$ as might be
expected; noise reduces the \apost probability of detection.

To illustrate this further, suppose again that $r=I_n/n$ and we use
the decomposition $\psi\psi^*=U\ \diag(0,\ldots,0,1)\ U^*$. This
gives,
\beq[eq:a1gam diag]
\Om_+ = (\gam^\opt_\noisy\bet/n)\ I_{n-1},
\;\;
\Om_- = \gam^\opt_\noisy\left( 1-\bet(1-1/n) \right) - (1-\bet)
\eeq
Consequently, \refeq{ocwc noise 1} holds if
\beq[eq:gamopt noise]
\gam^\opt_\noisy
=
\frac{1-\bet}
{
1-\bet\left(
1-\frac{1}{n}-\frac{\nu_0}{1-\nu_0}\frac{n-1}{n}
\right)
}
\eeq
This clearly shows that $\gam^\opt_\noisy <
\gam^\opt=(1-\bet)/(1-\bet(1-1/n))$. In addition ,as $n\to\infty$,
\beq[eq:gamopt noisy inf]
\gam^\opt_\noisy 
\to 
\frac{1-\bet}{1-\bet\left(\frac{1-2\nu_0}{1-\nu_0}\right)}
\eeq
When $\nu_0>1/2$, the matrix inequalities in \refeq{ocwc noise}
reverse and become,
\beq[oc matineq rev]
A_1(\gam^\opt_\noisy)O_2 \leq 0,
\;\;
A_1(\gam^\opt_\noisy)O_1 \geq 0
\eeq
These are satisfied if the POVM matrices for $\nu_0<1/2$ are also
reversed, that is, set to $O_1=U_+U_+^*,\ O_2=U_-U_-^*$ with
$(U_+,U_-)$ from the decomposition \refeq{uomu noise}. For a fixed
$\bet$, the optimal objective value with $\nu_0<1/2$ will always be
greater than the value with $\nu_0>1/2$. When $\nu_0=1/2$, this type
of detector can do no better than $\gam=1-\bet$, the occurrence
probability for the pure state.

%%%%%%%%%%%%%%%%%%%%%%%%
\subsection{Unambiguous detection}
\label{sec:umb}

As discussed briefly in Section \refsec{perfumb}, an unambiguous
detector is one that with some probability either detects the correct
state or else declares the result inconclusive. This requires an
additional POVM element to account for the inconclusive result.
Specifically, as before, let $O_i,\ i=1,\ldots,m$ correspond to the
$m$ input states $\rho_i,\ i=1,\ldots,m$ and let $O_0$ correspond to
the inconclusive result.  Thus there are $m+1$ POVM elements, $O_i,\
i=0,\ldots,m$. The probability of an inconclusive result is therefore,
\beq[eq:pincl 1]
\pincl = \trace\ O_0 \rho
\eeq
The ideal unambiguous detector is one where the \apost probability
error is zero, or equivalently $\pinout(i|i)=1,\ i=1,\ldots,m$ which
can only occur when \refeq{imperfect} holds which here becomes,
\beq[eq:umamb]
\poutin(i|j)
=
\trace\ O_i \rho_j
=
\pb(i)\ \del_{ij},\;
i,j=1,\ldots,m
\eeq
Observe that if $\pb(i)=1,\ i=1,\ldots,m$ then from \refeq{pincl 0}
$\pincl=0$, and hence, $O_0=0$, which eliminates the need for the
extra (inconclusive) detector outcome and reduces to the condition for
{\em perfect} detection \refeq{perfect}. Allowing for a non-zero
probability of an inconclusive result opens the possibility that a
detector can be designed to satisfy \refeq{unamb 1}, and thus,
$e_\post(i)=0,\ i=1,\ldots,m$. 

%%%%%%%%%%%%%%%%%%%%%%%%%%%%%%%%%
\subsubsection*{Optimal \apost performance with an inconclusive outcome}

By relaxing the requirement for zero error we can find a randomized
detector by solving the following problem.
\beq[eq:optumb]
\bea{ll}
\mbox{minimize}
&
\normwc{e_\post}
=
\ds
\max_{i=1,\ldots,m}\
w_i
\left(
1-
\frac{
p_i \ \trace\ O_i\ \rho_i
}{
\trace\ O_i \rho
}
\right)
\\
\mbox{subject to}
&
\sum_{i=0}^m\ O_i = I_n,
\;\;\;
O_i \geq 0,\;
i=0,\ldots,m
\eea
\eeq
If $\normwc{e_\post}=0$ then we have found an {\em unambiguous
detector}, one that either produces the correct result or is
inconclusive. Otherwise the detector is randomized, but not
unambiguous.  However, the extra design freedom in the inconclusive
POVM matrix, $O_0$, insures that the resulting $\normwc{e_\post}$ will
always be smaller than the value obtained for a detector without the
additional inconclusive outcome.
%%%%%%%%%%%%%%%%%%%%%%%%%%%%%%%%%%%%%%%%%%%
\subsubsection*{Optimality conditions}

Observe also that the only difference between problem \refeq{optumb}
and problem \refeq{optpost} {\em is} the extra POVM element $O_0$. As
a result the optimality conditions have extra constraints to account
for the additional element. Specifically, any feasible POVM is optimal
if and only if there exist real constants $\lam_i,\ i=1,\ldots,m$ such
that,
\beq[eq:oc umb]
\bea{rcl}
\lam_i &>& 0,\; i\in S
\\
\lam_i &=& 0,\; i\not\in S
\\
\sum_{i=1}^m\ \lam_i &=& 1
\\
\lam_i A_i(\del^\opt)
-\sum_{j=1}^m\ \lam_j A_j(\del^\opt) O_j &\geq& 0,\;
i=1,\ldots,m
\\
\left( 
\lam_i A_i(\del^\opt)
-\sum_{j=1}^m\ \lam_j A_j(\del^\opt)O_j
\right)
O_i &=& 0,\;
i=1,\ldots,m
\\
\sum_{j=1}^m\ \lam_j A_j(\del^\opt) O_j &\leq& 0
\\
\left(
\sum_{j=1}^m\ \lam_j A_j(\del^\opt) O_j
\right)
O_0 &=& 0
\eea
\eeq
The index set $S$ consists only of those indices where the optimal
objective value from \refeq{optumb} is achieved. Thus,
\beq[eq:idx post umb]
S = \set{i=1,\ldots,m}{\trace\ O_i A_i(\del^\opt) =0}
\eeq
If the optimal is achieved at only one constraint, say $i=k$, then
$\lam_k=1\, \lam_i=0,\ i\neq k$, and the optimality conditions become:
\beq[eq:ocwc umb noise]
\bea{rcl}
A_k(\del^\opt)(I-O_k) &\geq& 0
\\
A_k(\del^\opt)O_k &\leq& 0
\\
A_k(\del^\opt)(I-O_k) O_k &=& 0
\\
\left( A_k(\del^\opt)O_k \right) O_i &=& 0,\ 
i\in\seq{1,\ldots,m}\neq k
\\
A_k(\del^\opt)O_k O_0 &=& 0
\eea
\eeq
%

%%%%%%%%%%%%%%%%%%%%%%%%%%%
\subsubsection*{Optimal \apost performance with an inconclusive outcome
and measurement noise}

Problem \refeq{optumb} can be modified to account for measurement
noise.
\beq[eq:optumbnoise]
\bea{ll}
\mbox{minimize}
&
\normwc{e_\post}
=
\ds
\max_{i=1,\ldots,m}\
w_i
\left(
1-
\frac{
p_i \ \trace\ O_i^\noisy\ \rho_i
}{
\trace\ O_i^\noisy \rho
}
\right)
\\
\mbox{subject to}
&
O^\noisy_i 
= 
\sum_{j=0}^m \ 
\nu_{ij}\
O_{j},
\;\;
i=0,\ldots,m
\\
&
\sum_{i=0}^m\ O_i = I_n,
\;\;\;
O_i \geq 0,\;
i=0,\ldots,m
\eea
\eeq
In this case because of the noise, it is doubtful that an unambiguous
detector can be found. Nonetheless, the resulting randomized detector
will still outperform one without an inconclusive outcome.

%%%%%%%%%%%%%%%%%%%%%%%%%%%%%%%%%%%%
\subsection{Example}
\label{sec:example}

Consider the following two (pure) input states and corresponding
occurrence probabilities:
\beq[eq:2in]
\bea{ll}
\rho_1
=
\left[
\bea{c}
1/\sqrt{2} \\ 1/\sqrt{2}
\eea
\right]\
\left[
\bea{c}
1/\sqrt{2} \\ 1/\sqrt{2}
\eea
\right]^T
&
\rho_2
=
\left[
\bea{c}
1 \\ 0
\eea
\right]\
\left[
\bea{c}
1 \\ 0
\eea
\right]^T
\\
\pin(1) = 2/3
&
\pin(2) = 1/3
\eea
\eeq
Throughout this example we place equal weights on each state,
\beq[eq:wei]
[w_1,w_2] =[1,\ 1]
\eeq
Optimizing the worst-case \apost probability measure,
\refeq{optpost}, returns \apost probabilities\footnote{
All numerical results were obtained using {\sc sedumi}
\cite{Sturm:99}. The numbers shown are rounded to two significant
digits.  }
\beq[eq:wcdet1]
\bea{ll}
\pinout(1|1)=0.87
&
\pinout(2|2)=0.87
\eea
\eeq
and POVM matrices which are well approximated by the rank-one
projectors\footnote{
The positive semi-definite matrices returned by the convex program are
approximated by rank-one projectors using a singular value decomposition
only when the maximum singular value is much greater than all the
others.
}
\beq[eq:wcdet2]
\left\{\
\left[\bea{c} 0.53 \\  0.85 \eea\right],\
\left[\bea{c} -0.85 \\  0.53 \eea\right]\
\right\}
\eeq
Optimizing the worst-case \apost probability measure with the
additional inconclusive outcome, \refeq{optumb}, for the bound
$\pincl\leq 1$, returns an unambiguous detector with \apost and
inconclusive probabilities,
\beq[eq:wcran1]
\bea{lll}
\pinout(1|1)=1
&
\pinout(2|2)=1
&
\pincl = 0.75
\eea
\eeq
The associated POVM matrices (rank-one projectors) are
\beq[eq:wcran2]
\left\{\
\left[\bea{c}  0     \\   0.62 \eea\right],\
\left[\bea{c}  -0.62 \\   0.62 \eea\right],\
\left[\bea{c}  0.79  \\   0.49 \eea\right]\
\right\}
\eeq
This is an unambiguous detector which is perfectly correct 75\% of the
time.

Now we add 2\% noise and solve \refeq{optpost 1} with
\beq[eq:lamdet]
\nu
=
\left[
\bea{cc} 
0.98 & 0.02
\\ 
0.02 & 0.98
\eea 
\right]
\eeq
By comparison with \refeq{wcdet1}-\refeq{wcdet2} we now get
\beq[eq:wcdet1n]
\bea{ll}
\pinout(1|1)=0.86
&
\pinout(2|2)=0.86
\eea
\eeq
and similar POVM rank-one projectors
\beq[eq:wcdet2n]
\left\{\
\left[\bea{c} 0.55 \\  0.83 \eea\right],\
\left[\bea{c} -0.83 \\  0.55 \eea\right]\
\right\}
\eeq
Solving \refeq{optumbnoise} with a similar 2\% noise
\beq[eq:lamran]
\nu_\incl
=
\left[
\bea{ccc} 
0.98 & 0.01 & 0.01 
\\ 
0.01 & 0.98 & .01
\\
0.01 & 0.01 & 0.98
\eea 
\right]
\eeq
gives the probabilities,
\beq[eq:wcran1n]
\bea{lll}
\pinout(1|1)=0.96
&
\pinout(2|2)=0.96
&
\pincl = 0.76
\eea
\eeq
and POVM matrices which are well approximated by the rank-one
projectors,
\beq[eq:wcran2n]
\left\{\
\left[\bea{c}  0.04     \\   0.46 \eea\right],\
\left[\bea{c}  -0.68 \\   0.66 \eea\right],\
\left[\bea{c}  0.73  \\   0.59 \eea\right]\
\right\}
\eeq
This is no longer an unambiguous detector but rather a randomized
detector. For 76\% of the time an inconclusive result will occur. When
the detector declares either state 1 or state 2, the probability of
being correct is 96\% which is better than the deterministic detector
with probabilities of 86\%. If the situation is such that there is
little penalty in waiting, then a higher probability outcome is
promised by the randomized detector.

\begin{figure}[h]
\centering
\figsfile{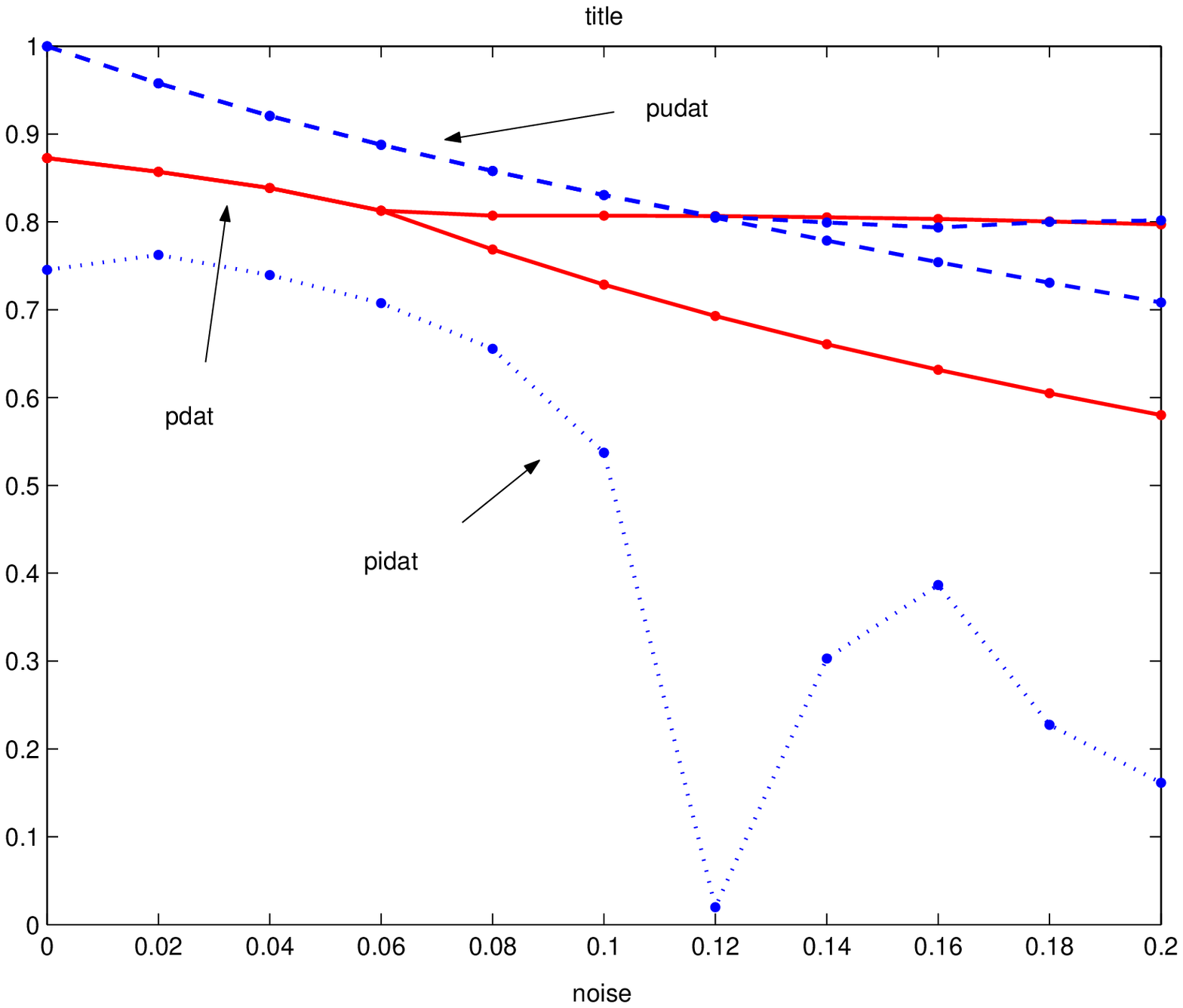,height=4in,width=6in}
\caption{$\pinout(i|i),\ i=1,2$ optimized for each noise level via
\refeq{optpost} and \refeq{optumb}.}
\label{fig:ppupidat}
\end{figure}

We now repeat all the above optimal designs for varying noise levels:
\beq[eq:lamgam]
\nu(\nu_0)
=
\left[
\bea{cc} 
1-\nu_0 & \nu_0
\\ 
\nu_0 & 1-\nu_0
\eea 
\right],\;
\nu_\incl(\nu_0)
=
\left[
\bea{ccc} 
1-\nu_0 & \nu_0/2 & \nu_0/2
\\ 
\nu_0/2 & 1-\nu_0 & \nu_0/2
\\
\nu_0/2 & \nu_0/2 & 1-\nu_0
\eea 
\right],\;
\nu_0\in[0,\ 0.20]
\eeq
The results are plotted in Figure \ref{fig:ppupidat} for $\nu_0$ from 0
to 0.20 in 0.02 increments. The solid curves are the two diagonal
elements of the \apost probability matrix for the optimal
randomized detector. Associated with them is the dotted curve showing
$\pincl$, the probability of an inconclusive result. The dashed curves
are the two diagonal elements of the \apost probability matrix for the
optimal deterministic detector. As expected, the randomized detector
outperforms the deterministic detector as seen by the fact that the
lower solid curve is always larger than the lower dashed curve. (The
optimal worst-case design maximizes the minimum error, which is
equivalent to making the lower of the two curves as large as
possible.) In all cases the POVMs were easily approximated by rank-one
projectors, but in no case were the projectors in the natural basis.

The behavior of $\pincl(\nu_0)$ is quite interesting. The inconclusive
probability and the associated POVM matrix become small at a noise
level $\eta \approx 0.12$, in effect, turning off the randomized
feature.

\begin{figure}[h]
\centering
\figsfile{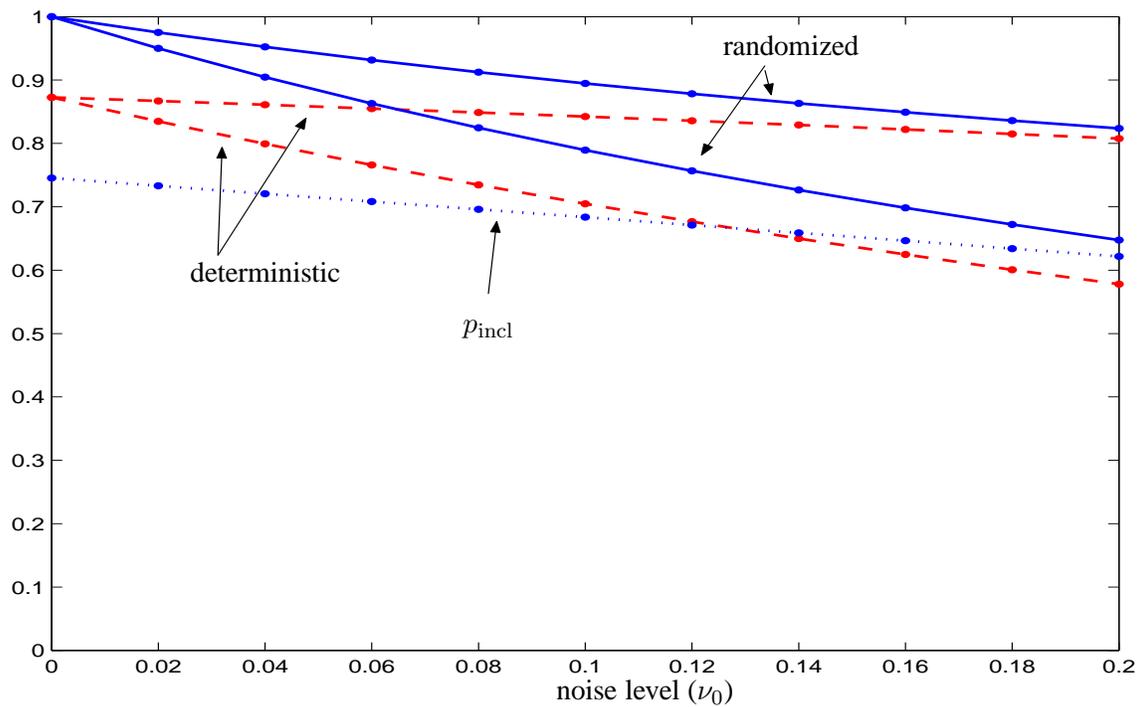,height=3.75in,width=6in}
\caption{$\pinout(i|i),\ i=1,2$ optimized only for the zero noise level.}
\label{fig:ppupi0dat}
\end{figure}

Figure \ref{fig:ppupi0dat} shows the robustness properties of the
randomized and deterministic detectors. We fixed the POVMs for the two
cases at their optimal settings corresponding to the noise-free case
$(\nu_0=0$). The plots show what happens as the noise level
increases. The probability levels are not all that different from the
optimal noisy results in Figure \ref{fig:ppupidat}, but are of course
not as good.

%%%%%%%%%%%%%%%
\psfrag{title}{$\pinout(i|i),\ i=1,2$}
\psfrag{noise}{noise level ($\nu_0$)}
\psfrag{pdat }{deterministic}
\psfrag{pudat }{randomized}
\psfrag{pidat }{$\pincl$}

%%%%%%%%%%%%%%%%%%%%%%%%

%%%%%%%%%%%%%%%
\psfrag{title}{$\pinout(i|i),\ i=1,2$}
\psfrag{noise}{noise level ($\nu_0$)}
\psfrag{pdat }{deterministic}
\psfrag{pudat }{randomized}
\psfrag{pidat }{$\pincl$}

%%%%%%%%%%%%%%%%%%%%%%%%
\clearpage

%%%%%%%%%%%%%%%%
\iffalse

%zdat=num2str([dndat,pdat,pudat,pidat],2)

   0     0.87     0.87        1        1     0.75\\
0.02     0.86     0.86     0.96     0.96     0.75\\
0.04     0.84     0.84     0.92     0.92     0.72\\
0.06     0.81     0.81     0.89     0.89     0.67\\
0.08     0.81     0.77     0.86     0.86     0.62\\
0.10     0.81     0.73     0.83     0.83     0.51\\
0.12     0.81     0.69     0.81     0.80     0.08\\
0.14     0.81     0.66     0.80     0.78     0.31\\
0.16     0.80     0.63     0.79     0.75     0.37\\
0.18     0.80     0.60     0.80     0.73     0.26\\
0.20     0.80     0.58     0.80     0.71     0.21\\

%z0dat=num2str([dndat,p0dat,pu0dat,pi0dat],2)

   0      0.87     0.87        1        1     0.75\\
0.02     0.87     0.83     0.98     0.95     0.73\\
0.04     0.86     0.80     0.95     0.90     0.72\\
0.06     0.85     0.77     0.93     0.86     0.71\\
0.08     0.85     0.73     0.91     0.82     0.70\\
0.10     0.84     0.70     0.89     0.79     0.68\\
0.12     0.84     0.68     0.88     0.76     0.67\\
0.14     0.83     0.65     0.86     0.73     0.66\\
0.16     0.82     0.62     0.85     0.70     0.65\\
0.18     0.82     0.60     0.84     0.67     0.63\\
0.20     0.81     0.58     0.82     0.65     0.62\\

\fi
%%%%%%%%%%%%%%%%%%%%%%%%%%%%%%%%%%%%%%%
\section{Extensions and Other Considerations}
\label{sec:future}

%%%%%%%%%%%%%%%%%%%%%%%%%%%%%%%%%%%%%%%%%%%%%%%%%%%%
\subsection{Uncertain dynamics}
\label{sec:unc dyn}

\begin{figure}[h]
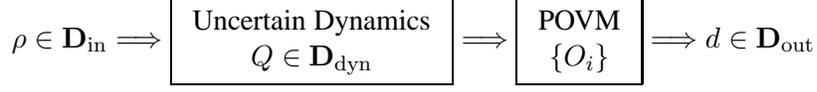

\[
\rho\in\Dbf_{\rm in}
\Longrightarrow
\mathbox{
\bea{c}
\mbox{Uncertain Dynamics}
\\
Q \in \Dbf_\dyn
\eea
}
\Longrightarrow
\mathbox{
\bea{c}
\mbox{POVM}
\\
\seq{O_i}
\eea
}
\Longrightarrow
d\in\Dbf_{\rm out}
\]
\caption{Detector with uncertain dynamics.}
\label{fig:detect uncdyn}
\end{figure}

The goal is to design the POVM $\seq{O_i}$ in the presence of
uncertain detector dynamics $Q\in\Dbf_\dyn$ as illustrated in Figure
\ref{fig:detect uncdyn}.

\noindent
We will assume that $\Dbf_\dyn$ consists of a finite number of unitary
operators $\seq{U_k}$ with corresponding occurrence probabilities
$\seq{\pdyn(k)}$. Thus,
\beq[eq:omdyn]
\bea{rcl}
\Dbf_\dyn
&=&
\set{U_k \in \Cbfnn}{k=1,\ldots,\ell}
\\
p_\dyn(k)
&=&
\prob{Q=U_k}
\eea
\eeq
The conditional probability \refeq{cond 1} now becomes,
\beq[eq:poutin dyn]
\poutin(i|j)
=
\trace\ O_i \rhoh_j\ ,
\;\;\;\;
\rhoh_j
=
\sum_{k=1}^\ell\
\pdyn(k) U_k \rho_j U_k^*
\eeq
This clearly shows that the only changes to make is to replace
$\rho_j$ with $\rhoh_j$ everywhere, specifically, in the error
probabilities \refeq{eprob 2} and in the output state $\rho$ as
defined by \refeq{rho}.

The above representation of $Q$ is an example of a the more generic
{\em Kraus operator sum representation} (OSR).  Specifically, the {\em
Kraus matrices}, $\set{K_k\in\Cbfnn}{k=1,\ldots,\ell}$ with $\ell \leq
n^2$, can characterize a large class of possibilities for the
$\qsys$-system as follows:
\beq[eq:osr]
\qsys(\rho,K)
=
\sum_{k=1}^\ell\ K_k \rho K_k^*,
\;\;\;
\sum_{k=1}^\ell\ K_k^* K_k = K_0 \leq I_n
\eeq
Comparing this with \refeq{omdyn} gives $K_k=\sqrt{\pdyn(k)} U_k$ and
$K_0=I_n$, which clearly is just one possibility.  For example, when
$K_0 < I_n$, additional measurement operations within $Q$ are
included. The OSR also accounts for many forms of error sources as
well as decoherence, \eg, \cite{NielsenC:00}, \cite{LidarETAL:01}.

%%%%%%%%%%%%%%%%%%%%%%%%%%%%%%%%
\subsection{Detector with fixed POVM}
\label{sec:fixed}

In this section we consider designing the detector for a fixed POVM
set. We will show that the detector dynamics when represented as an
OSR (Operator-Sum-Representation) can also be designed by solving a
quasiconvex optimization problem.

\begin{figure}[h]
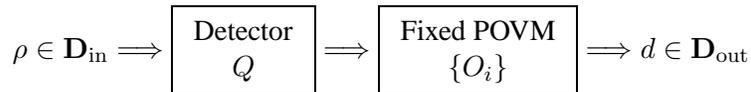

\[
\rho\in\Dbf_{\rm in}
\Longrightarrow
\mathbox{
\bea{c}
\mbox{Detector}
\\
Q
\eea
}
\Longrightarrow
\mathbox{
\bea{c}
\mbox{Fixed POVM}
\\
\seq{O_i}
\eea
}
\Longrightarrow
d\in\Dbf_{\rm out}
\]
\caption{Detector with fixed POVM.}
\label{fig:detect fixpovm}
\end{figure}

Suppose we are {\em given} the POVM, $\seq{O_i}$, and wish to design $Q$
for optimal detection as shown in Figure \ref{fig:detect fixpovm}.  

\noindent
The POVM would be most likely selected as rank-one projectors in the
natural basis.  For example, for $i=1,\ldots,m$, fix $b_i\in\Cbf^m$
and $O_i = I_\ell \otimes b_i\ b_i^*$ with $\ell+m=n$, the dimension
of the input state. The input state might also consist of prepared
ancilla states.  In the natural basis $b_1^T=[1\ 0\ \cdots\ 0],\
b_2=[0\ 1\ 0\ \cdots\ 0],\ldots,b_m^T=[0\ \cdots\ 0\ 1]$.

As noted in Section \refsec{unc dyn}, a very general form to
characterize $Q$ is the Krause OSR. Using \refeq{osr}, the \apost
performance probability is now,
\beq[eq:postq]
\ppost(i)
=
\frac{
p_i \ \trace\ O_i\ Q(\rho_i,K)
}{
\trace\ O_i Q(\rho,K)
}
\eeq
which is quadratic (fractional) in the Kraus matrices. It can be
transformed into a quasiconvex function by expanding the Kraus
matrices in a fixed basis. The procedure, described in
\cite[\S 8.4.2]{NielsenC:00}, is as follows: since any matrix in
$\Cbfnn$ can be represented by $n^2$ complex numbers, let
\beq[eq:basis]
\set{B_\mu\in\Cbfnn}{\mu=1,\ldots,n^2}
\eeq
be a basis for matrices in $\Cbfnn$. The Kraus matrices can thus be
expressed as,
\beq[eq:krausb]
K_k = \sum_{\mu=1}^{n^2}\ a_{k\mu} B_\mu,\;
k=1,\ldots,\ell
\eeq
where the $n^2$ coefficients $\seq{a_{k\mu}}$ are complex scalars. As
shown in \cite{NielsenC:00} the representation \refeq{osr} now
becomes,
\beq[eq:osr1]
\qsys(\rho,X)
=
\sum_{\mu,\nu=1}^{n^2}\ X_{\mu\nu}\ B_\mu \rho B_\nu^*,
\;\;\;
\sum_{\mu,\nu=1}^{n^2}\ X_{\mu\nu}\ B_\mu^* B_\nu \leq I_n
\eeq
with
\beq[eq:osr2]
X_{\mu\nu} = \sum_{k=1}^\ell\ a_{k\mu}^* a_{k\nu},
\;\;\;
\mu,\nu=1,\ldots,n^2
\eeq
The matrix $X\in\Cbf^{n^2\times n^2}$ with the above coefficients must
also be non-negative in order to maintain probabilities. The number
of free (real) variables in $X$ is thus $n^4-n^2$. In addition, we can
write,
\beq[eq:trosr]
\pout(i)
=
\trace\ O_i\ Q(\rho,X)
=
\trace\ X R_i(\rho),\;
i=1,\ldots,m
\eeq
where the matrix $R_i(\rho)\in\Cbf^{n^2\times n^2}$ has elements
given by,
\beq[eq:rmat]
[R_i(\rho)]_{\mu\nu}
=
\trace\ B_\nu \rho B_\mu^* O_i,\;
\mu,\nu = 1,\ldots,n^2
\eeq
The problem of optimally designing the ``system'' part of the
detector, the $\qsys$-system, is equivalent to the following
optimization problem over the positive semidefinite matrix
$X\in\Cbf^{n^2\times n^2}$.
\beq[eq:optpostq]
\bea{ll}
\mbox{minimize}
&
\normwc{e_\post}
=
\ds
\max_{i=1,\ldots,m}\
w_i\
\left(
1-
\frac{
p_i \ \trace\ X R_i(\rho_i)
}{
\trace\ X R_{i}(\rho)
}
\right)
\\
\mbox{subject to}
&
\sum_{\mu\nu}\ X_{\mu\nu}\ B_\mu^*B_\nu \leq I_n,
\;\;\;
X \geq 0
\eea
\eeq
This problem, like \refeq{optpost}, is also a quasiconvex optimization
problem with the optimization variables being the elements of the
matrix $X$.

%%%%%%%%%%%%%%%%%%%%%%%%%%%%%%%%%%%%
\subsubsection*{Implementation of OSR}
%\label{sec:imp osr}

An OSR can be implemented using unitary operations (and if necessary
projection measurements) and the $X$-matrix can be transformed to Kraus
operators via the singular value decomposition \cite{NielsenC:00}.
Specifically, let $X=VSV^*$ with unitary $V\in \Cbf^{n^2\times n^2}$
and $S={\rm diag}(s_1\ \cdots\ s_{n^2})$ with the singular values
ordered so that $s_1 \geq s_2 \geq\ \cdots\ \geq s_{n^2} \geq 0$. Then
the coefficients in the basis representation of the Kraus matrices
\refeq{krausb} are,
\beq[eq:aki]
a_{k\mu} = \sqrt{s_k}\ V_{\mu k}^*,\
k,\mu=1,\ldots,n^2
\eeq
Theoretically there can be fewer then $n^2$ Kraus operators. For
example, if the $\qsys$ system is unitary, then,
\beq[eq:u rho u]
\qsys(\rho) = U \rho U^*
\eeq
In effect, there is one Kraus operator, $U$, which is unitary and of
the same dimension as the input state $\rho$.  The corresponding $X$
matrix is a dyad, hence $\rank\ X=1$. Adding a rank constraint would
thus force a simplification of the implementation. Unfortunately, a
rank constraint is not convex. However, the $X$ matrix is symmetric
and positive semidefinite, hence the heuristic from \cite{FazelHB:01}
applies where the rank constraint is replaced by the trace constraint,
\beq[eq:rankx]
\trace\ X \leq \eta
\eeq
From the singular value decomposition of $X$, $\trace\ X = \sum_k
s_k$. Adding the constraint \refeq{rankx} to \refeq{optpostq} will
force some (or many) of the $s_k$ to be small which can be eliminated
(post-optimization) thereby reducing the rank. The auxiliary parameter
$\eta$ can be used to find a tradeoff between simpler realizations and
performance.

%%%%%%%%%%%%%%%%%%%%%%%%%%%
%\section{Concluding remarks}

%%%%%%%%%%%%%%%%%%%%%%%%%%%%%%%%%%%
\section*{Acknowledgements}

This work was supported by the DARPA QuIST Program ({\bf Qu}antum {\bf
I}nformation {\bf S}cience \& {\bf T}echnology). The first author is
grateful for numerous discussions with Stephen Boyd of Stanford
University on convex optimization and with Abbas Emami-Naeini of SC
Solutions on matrix analysis.

%%%%%%%%%%%%%%%%%%%%%%%%%%%%%%%%
%\newpage
\appendix
\section{Optimality Conditions}
\label{sec:cvxopt}

Optimality conditions are derived from Lagrange Duality Theory for the
following detection criteria: (i) average joint performance, (ii)
worst-case \apost performance with noise-free measurements, and (iii)
worst-case \apost performance with noisy measurements.

{\bf Caveat emptor} The material in this section is meant to be a
``scaffold'' to what can be found in some of the recent texts on
convex optimization, \eg, see \cite{BoydV:04} and the references
therein. More specifically, we refer principally to the sections in
\cite{BoydV:04} where detailed information and proofs can be found for
any axiomatic statements made here. The same caution applies to our
references to computational methods: interested readers should refer
directly to the available convex solvers which can be downloaded from
the web, \eg, {\sc sdpsol} \cite{WuB:00} or {\sc sedumi}
\cite{Sturm:99}.

%%%%%%%%%%%%%%%%%%%%%%%%%%%%%%%%%%%
\subsection{Optimality conditions for average joint performance}
\label{sec:oc joint}

We will apply Lagrange Duality Theory \cite[Ch.5]{BoydV:04} to the
optimization problem \refeq{optjoint av} referred to in this context
as the {\em primal problem}. The Lagrange function associated with the
primal problem \refeq{optjoint av} is,
\beq[eq:lag joint]
L(O,Z,Y)
=
\sum_{i=1}^m\
\trace\ O_i A_i-\trace\ Z_i O_i
+
\trace\ Y\Big( I_n - \sum_{i=1}^m\ O_i \Big)
\eeq
with Lagrange multipliers $Z_i\in\Cbfnn,\ Z_i \geq 0$ for the
inequality constraint $O_i\geq 0$, and $Y\in\Rbf^{n\times n},\ Y=Y^T$
for the equality constraint $\sum_{i=1}^m\ O_i = I_n$.  The first term
in $L(O,Z,Y)$ is the objective function in \refeq{optjoint av}
expressed in terms of the data matrices $A_i$ from \refeq{optavjoint
amat}. The {\em Lagrange dual function} is defined as,,
\beq[eq:dual fun joint]
\bea{rcl}
g(Z,Y)
&=&
\ds
\inf_{O}\
L(O,Z,Y)
\\
&=&
\left\{
\bea{ll}
\trace\ Y
&
A_i-Z_i-Y=0,\
i=1,\ldots,m
\\
-\infty
&
\mbox{otherwise}
\eea
\right.
\eea
\eeq
One of the important properties of the dual function is that for any
$Z_i\geq 0$ and any $Y$, we get the lower bound,
\beq[eq:lowbnd joint]
g(Z,Y) \leq \del^\opt
\eeq
where $\del^\opt$ is the optimal objective value from solving
\refeq{optjoint av}.  The {\em Lagrange dual problem} establishes the
largest lower bound from,
\beq[eq:dual pblm joint]
\bea{ll}
\mbox{maximize}
&
g(Z,Y)
\\
\mbox{subject to}
&
Z_i \geq 0,\;
i=1,\ldots,m
\eea
\eeq
where the optimization variables are $(Z,Y)$. Using \refeq{dual fun
joint} we can eliminate the $Z_i$ variables and write the dual problem
explicitly in terms of the $Y$ variables as,
\beq[eq:dual pblm 1 joint]
\bea{ll}
\mbox{maximize}
&
\trace\ Y
\\
\mbox{subject to}
&
A_i - Y \geq 0,\;
i=1,\ldots,m
\eea
\eeq
A solution, $Y^\opt$, the {\em dual optimal multiplier}, also
returns the maximum objective value, $d^\opt=\trace\ Y^\opt$, the {\em
dual optimal value}. From \refeq{lowbnd joint} we get $d^\opt \leq
\del^\opt$. A numerical solution of the primal problem \refeq{optjoint
av} always returns $\delh \geq \del^\opt$, and likewise numerically
solving the dual problem \refeq{dual pblm 1 joint} will always return
$\dh \leq d^\opt$. Thus, the optimal solution is always contained in
the known interval
$
\dh 
\leq 
d^\opt
\leq 
s^\opt
\leq
\sh
$
. For this primal-dual pair we also have {\em strong duality}, that
is, $d^\opt=s^\opt$. This follows because the primal problem satisfies
{\em Slater's condition} \cite[\S 5.2.3]{BoydV:04}, which in this case
means that the primal problem is convex and there exist strictly
feasible $(O_i)$, \ie, $O_i > 0,\; i=1,\ldots,m,\; \sum_{i=1}^m\ O_i =
I_n $. (For example, let $O_i=I_n/m$). The optimal and computed
objective values then satisfy,
\beq[eq:bnds joint] 
\dh 
\leq 
\trace\ Y^\opt
=
\del^\opt
\leq
\delh
\eeq
Strong duality also implies the following {\em complementary
slackness} conditions, \cite[\S 5.5.2]{BoydV:04},
\beq[eq:slack joint]
\bea{rcl}
Z^\opt_i O^\opt_i 
= 
\left( A_i -Y^\opt \right) O^\opt_i 
&=&
0,\;
i=1,\ldots,m
\eea
\eeq
The last line uses $Z_i=A_i-Y$ from \refeq{dual fun joint}.
Combining $\sum_{i=1}^m\ O^\opt_i =I$ with $(A_i-Y^\opt ) O^\opt_i =
0,\; i=1,\ldots,m$ gives,
\beq[eq:yopt joint]
Y^\opt = \sum_{i=1}^m\ A_i O^\opt_i
\eeq
This can be used in to eliminate $Y^\opt$ in \refeq{dual pblm 1 joint}
and \refeq{slack joint} yielding the constraints,
\beq[eq:oc joint]
\bea{rcl}
A_i - \sum_{j=1}^m\ A_j O_j &\geq& 0,\;
i=1,\ldots,m
\\
\left( A_i - \sum_{j=1}^m\ A_j O_j \right) O_i &=& 0,\;
i=1,\ldots,m
\eea
\eeq
These are the conditions stated in \refeq{optavjoint cond} as being
necessary and sufficient for optimality of any feasible POVM set
$\seq{O_i}$. The proof of this statement relies on the fact that if
strong duality holds and the primal problem is convex -- both true for
this problem -- then the above conditions \refeq{oc joint} are
equivalent to the Karush-Kuhn-Tucker (KKT) conditions for optimality,
which in this case are both necessary and sufficient \cite[\S
5.5.3]{BoydV:04}. Thus, {\em any} feasible POVM set which satisfies
\refeq{oc joint} is optimal.

%%%%%%%%%%%%%%%%%%%%%%%%%%%%%%%%%%%%%%%%%%%%%%%%%%%%%%%%
\subsection{Optimality conditions for worst-case \apost performance}
\label{sec:oc apost}

As shown in \cite[\S 4.2.5]{BoydV:04}, a solution to the quasiconvex
optimization problem \refeq{optpost} can be obtained by solving a
series of convex feasibility problems together with a bisection
method.  We start with the equivalence,
\beq[eq:del]
\normwc{e_\post} \leq \del
\Leftrightarrow
w_i
\left(
1-
\frac{
p_i \ \trace\ O_i\ \rho_i
}{
\trace\ O_i \rho
}
\right)
\leq \del
\Leftrightarrow
\left\{
\bea{l}
\trace\ O_i A_i(\del) \leq 0
\\
\\
A_i(\del) = (w_i-\del)\rho - w_i p_i  \rho_i
\eea
\right.
\eeq
Problem \refeq{optpost} is then equivalent to,
\beq[eq:optpost 2]
\bea{ll}
\mbox{minimize}
&
\del
\\
\mbox{subject to}
&
\trace\ O_i A_i(\del) \leq 0
\\
&
\sum_{i=1}^m\ O_i = I_n,
\;\;\;
O_i \geq 0,\;
i=1,\ldots,m
\eea
\eeq
where the variables are now the real scalar $\del$ as well as the POVM
matrices $\seq{O_i\in\Cbfnn}$. The algorithm below requires knowing an
upper and lower bound on the optimal $\del^\opt$. Without loss of
generality we can normalize the weights so that $0\leq w_i \leq 1$.
Since the objective is a weighted error probability, the feasible
range is $0 \leq \del^\opt \leq 1$. The bisection algorithm as
presented in \cite[\S 4.2.5]{BoydV:04} now becomes:

\bquote{\bf Bisection-Feasibility Method}
\blist
\item {\bf given} $\del_{\min}=0,\  \del_{\max}=1$, 
tolerance $\epsilon > 0$.

\item {\bf repeat}

\ben
\item $\del = (\del_{\min}+\del_{\max})/2$

\item Solve the convex feasibility problem
\beq[eq:cvxfeas]
\bea{ll}
\mbox{find} & O_i,\; i=1,\ldots,m
\\
\mbox{subject to}
&
\trace\ O_i A_i(\del) \leq 0
\\
&
\sum_{i=1}^m\ O_i = I_n,
\;\;\;
O_i \geq 0,\;
i=1,\ldots,m
\eea
\eeq

\item if feasible, $\del_{\max} = \del$; else $\del_{\min}=\del$
\een

\item {\bf until} $\del_{\max}-\del_{\min} \leq \epsilon$.
\elist

\equote
The feasibility step is equivalent to solving the following SDP in the
variables $(s,O_i)$:
\beq[eq:feas s]
\bea{ll}
\mbox{minimize} & s
\\
\mbox{subject to}
&
\trace\ O_i A_i(\del) \leq s
\\
&
\sum_{i=1}^m\ O_i = I_n,
\;\;\;
O_i \geq 0,\;
i=1,\ldots,m
\eea
\eeq
Let $s^\opt,\ O^\opt_i$ denote the optimal solution.
Under the temporary assumption that $\trace\ O_i^\opt \rho>0$,
the inequality $\trace\ O_i^\opt A_i(\del) \leq s^\opt$ is
equivalent to,
\beq[eq:epostdel]
\normwc{e_\post}
=\max_i\
w_i
\left(
1-\frac{p_i  \trace\ O_i^\opt \rho_i}
{\trace\ O_i^\opt \rho}
\right)
\leq
\del
+\frac{s^\opt}{\min_i\ \trace\ O_i^\opt \rho}
\eeq
It follows that if $s^\opt > 0$ then $\del$ is feasible, and hence,
$\del^\opt < \del$.  If $s^\opt < 0$ then $\del$ is infeasible, \ie,
$\del^\opt > \del$. The optimal value $\del^\opt$ is clearly the
solution to $s^\opt(\del^\opt)=0$. The bisection algorithm together
with using an interior-point method to solve the SDP \refeq{feas s}
will return a value of $\del$ to within any desired, but finite,
accuracy of the optimal. 

The key computational step is solving the feasibility problem
\refeq{feas s}.  High quality code which uses an interior-point method
is recommended such as those found in {\sc sdpsol} \cite{WuB:00} or
{\sc sedumi} \cite{Sturm:99}.  In many cases the optimal POVM matrices
are rank deficient which may result in a large condition number in the
linear equations to be solved in the Newton step. This should not be a
problem for well conceived code.

To obtain the optimality conditions we will now apply Lagrange Duality
Theory to the feasibility problem \refeq{feas s} in the
Bisection-Feasibility method. Problem \refeq{feas s} is the primal
problem.  As previously noted, the primal optimal value, $s^\opt(\del)$,
determines if $\del$ is feasible, Specifically,
\beq[eq:primalopt]
\bea{rcl}
s^\opt < 0 &\Leftrightarrow& \del > \del^\opt 
\\
s^\opt > 0 &\Leftrightarrow& \del < \del^\opt
\\
s^\opt = 0 &\Leftrightarrow& \del = \del^\opt
\eea
\eeq
The Lagrange function associated with the primal problem \refeq{feas
s} is,
\beq[eq:lag]
L(s,O,\lam,Z,Y)
=
s+
\sum_{i=1}^m\
\Big(
\lam_i (\trace\ O_i A_i(\del)-s)-\trace\ Z_i O_i
\Big)
+
\trace\ Y\Big( I_n - \sum_{i=1}^m\ O_i \Big)
\eeq
with Lagrange multipliers $\lam_i\in\Rbf,\ \lam_i\geq 0$ for the
inequality constraint $\trace\ O_i A_i(\del) \leq s$, $Z_i\in\Cbfnn,\
Z_i \geq 0$ for the inequality constraint $O_i\geq 0$, and
$Y\in\Rbf^{n\times n},\ Y=Y^T$ for the equality constraint $\sum_{i=1}^m\
O_i = I_n$. The Lagrange dual function is then,
\beq[eq:dual fun]
\bea{rcl}
g(\lam,Z,Y)
&=&
\ds
\inf_{s,O}\
L(s,O,\lam,Z,Y)
\\
&=&
\left\{
\bea{ll}
\trace\ Y
&
\sum_{i=1}^m\ \lam_i=1,\;
\lam_i A_i(\del)-Z_i-Y=0,\
i=1,\ldots,m
\\
-\infty
&
\mbox{otherwise}
\eea
\right.
\eea
\eeq
The Lagrange dual problem establishes the largest lower bound from,
\beq[eq:dual pblm]
\bea{ll}
\mbox{maximize}
&
g(\lam,Z,Y)
\\
\mbox{subject to}
&
\lam_i \geq 0,\;
Z_i \geq 0,\;
i=1,\ldots,m
\eea
\eeq
where the optimization variables are $(\lam,Z,Y)$. Using \refeq{dual
fun} we can eliminate the $Z_i$ variables and write the dual problem
explicitly in terms of the $\lam_i$ and $Y$ variables as,
\beq[eq:dual pblm 1]
\bea{ll}
\mbox{maximize}
&
\trace\ Y
\\
\mbox{subject to}
&
\lam_i \geq 0,\;
\lam_i A_i(\del) - Y \geq 0,\;
i=1,\ldots,m
\\
&
\sum_{i=1}^m\ \lam_i =1
\eea
\eeq
The dual optimal solution is $(\lam^\opt,\ Y^\opt)$. Strong duality
also holds for this problem because Slater's condition holds \cite[\S
5.2.3]{BoydV:04}: there exist strictly feasible $(s,\ O_i)$, such that
$\trace\ O_i A_i(\del) < s,\; O_i > 0,\; i=1,\ldots,m,\; \sum_{i=1}^m\
O_i = I_n $. Since the primal (feasibility) problem is convex, the
optimal primal and dual objective values are equal,
\beq[eq:bnds] 
\trace\ Y^\opt
=
s^\opt
\eeq
Strong duality also implies the following complementary slackness
conditions, \cite[\S 5.5.2]{BoydV:04},
\beq[eq:slack]
\bea{rcl}
\lam^\opt_i\ 
\left(
\trace\ O^\opt_i A_i(\del) - s^\opt
\right)
&=&
0,\;
i=1,\ldots,m
\\
Z^\opt_i O^\opt_i 
= 
\left( \lam_i^\opt A_i(\del) -Y^\opt \right) O^\opt_i 
&=&
0,\;
i=1,\ldots,m
\eea
\eeq
The last line uses $Z_i=\lam_i A_i(\del) -Y$ from \refeq{dual fun}.
Combining $\sum_{i=1}^m\ O^\opt_i =I$ with $( \lam^\opt_i
A_i(\del^\opt)-Y^\opt ) O^\opt_i = 0,\; i=1,\ldots,m$ gives,
\beq[eq:yopt]
Y^\opt = \sum_{i=1}^m\ \lam^\opt_i A_i(\del) O^\opt_i
\eeq
We now put all the primal and dual equality and inequality constraints
together at the optimal $\del=\del^\opt$, $s^\opt=\trace\ Y^\opt =0$,
and use \refeq{yopt} to eliminate $Y^\opt$. To simplify notation we
drop the superscript $(\cdot)^\opt$ from all the variables
$(O,\lam,Y,Z,\del)$.  This gives:
\beq[eq:kkt]
\bea{rcl}
\sum_{i=1}^m\ O_i &=& I
\\
O_i &\geq& 0,\;
i=1,\ldots,m
\\
\lam_i \trace\ O_i A_i(\del) &=& 0,\;
i=1,\ldots,m
\\
\lam_i A_i(\del)-\sum_{j=1}^m\ \lam_j A_j(\del) O_j &\geq& 0,\;
i=1,\ldots,m
\\
\left( 
\lam_i A_i(\del)-\sum_{j=1}^m\ \lam_j A_j(\del) O_j 
\right) O_i  &=& 0,\;
i=1,\ldots,m
\\
\lam_i &\geq& 0,\;
i=1,\ldots,m
\\
\sum_{i=1}^m\ \lam_i &=& 1
\eea
\eeq
These can also be established directly from the KKT conditions for
optimality which in this case are both necessary and sufficient
\cite[\S 5.5.3]{BoydV:04}. For the linear constraints, either the
constraint is active, $\trace\ A_i(\del)O_i=0,\ \lam_i>0$, or
inactive, $\trace\ A_i(\del)O_i<0,\ \lam_i=0$. Combining this
with \refeq{kkt} gives the optimality conditions in \refeq{optwcpost
cond}.

Suppose the weights are all equal with $w_i=1,\ \forall i$.  Then,
\beq
A_i(\del) = \gam\rho - p_i  \rho_i \equiv A_i(\gam)
\eeq
with $\gam=1-\del$. From now on we will use $A_i(\gam)$ or $A_i(\del)$
as appropriate to the context.

Suppose the optimal is achieved by only one constraint, that is, for
$i=k$, $\trace\ O_k A_k(\gam) = 0$ and for $i\neq k$, $\trace\
O_i A_i(\gam) < 0$. Then, $\lam_k=1$, $\lam_{i\neq k}=0$ and the
optimality conditions \refeq{kkt} reduce to,
\beq[eq:optwcpost cond 1]
\bea{rcl}
A_k(\gam) \left(I-O_k\right) &\geq& 0
\\
A_k(\gam) O_k &\leq& 0
\\
A_k(\gam) \left(I-O_k\right) O_k &=& 0
\\
A_k(\gam) O_k O_i &=& 0,\;
i\neq k
\\
\trace\ O_k A_k(\gam) &=& 0
\eea
\eeq
Since $\rho>0$ by assumption \refeq{rhopos},
\[
\bea{rcl}
\det\ A_k(\gam)
&=&
\det\ 
\left(
\rho^{1/2}\left(
\gam I - p_k \rho^{-1/2}\rho_k\rho^{-1/2}
\right)
\rho^{1/2}
\right)
\\
&=&
(\det\ \rho)\ \prod_{j=1}^n\ \left( \gam - p_k \om_{kj} \right)
\eea
\]
with $\om_{kj},\ j=1,\ldots,n$ the eigenvalues of
$\rho^{-1/2}\rho_k\rho^{-1/2}$. Because $\rho>0,\ \rho_k\geq 0$, they
are all non-negative and $\max_j\om_{kj}>0$. Let
$\gam=p_k\max_j\om_{kj}$, or equivalently, 
\beq[eq:optgam]
\gam=p_k \ \sigmax(\rho^{-1/2} \rho_k \rho^{-1/2}) 
\eeq
where $\sigmax(\cdot)$ is the maximum singular value of the matrix
argument. With this choice $\det A_k(\gam)=0$ and hence $A_k(\gam)=0$
has the decomposition:
\beq[eq:svd akgam]
\bea{l}
A_k(\gam)
=
\left[U_{k+}\ U_{k0} \right]\
\left[
\bea{cc} \Om_{k+} & 0_{n-1} 
\\ 
0_{n-1}^T & 0 \eea
\right]\
\left[
\bea{c} U_{k+}^* \\ U_{k0}^* \eea
\right]
\\
\\
\Om_{k+} = {\bf diag}(\om_1,\ldots,\om_{n-1}),
\;
\om_1 \geq \om_2 \geq \cdots \geq \om_{n-1} \geq 0,
\;\; \om_1>0
\eea
\eeq
for unitary $[U_{k+}\ U_{k0}]$ with $U_{k+}\in\Cbf^{n\times n-1}$ and
$U_{k0}\in\Cbf^{n\times 1}$. Setting,
\beq[eq:povm akgam]
O_k = U_{k0} U_{k0}^*,
\;\;
I-O_k = U_{k+} U_{k+}^*
\eeq
gives $A_k(\gam)O_k=U_{k+}\Om_{k+}U_{k+}^*\geq 0,\
A_k(\gam)(I-O_k)=0,\ O_k(I-O_k)=0$, thus satisfying the optimality
conditions.  Observe also that $O_k$ is a rank $1$ projector, and
$I-O_k$ is a rank $n-1$ projector. Also, $I-O_k=\sum_{i\neq k} O_i$,
and hence is the sum of the remaining $m-1$ POVM elements. These are
thus arbitrary except for satisfying \refeq{povm akgam} with each
$O_i\geq 0,\ i\neq k$.

Since the single active constraint $k$ can occur for any
$i=1,\ldots,m$, then, 
\beq[eq:gam ak]
\gam = \min_{i=1,\ldots,m}\
p_i \sigmax(\rho^{-1/2} \rho_i \rho^{-1/2}) 
\eeq
which establishes \refeq{optval} as the optimal objective value for
equal weights with one active linear constraint. More specifically,
this means that there is a single index $k\in{1\ldots,m}$ such that
$\gam=p_k \sigmax(\rho^{-1/2} \rho_k \rho^{-1/2}) < p_i
\sigmax(\rho^{-1/2} \rho_i \rho^{-1/2}),\; \forall i\neq k$. 

The same procedure involving the decomposition \refeq{svd akgam} is
used to arrive at the results for single pure state detection with
weights $w_1=1,\ w_2=0$ given by \refeq{ocwc}-\refeq{povm pure}.

%%%%%%%%%%%%%%%%%%%%%%%%%%%%%%%%%%%%%%%%%%%%%%%%%%%%%%%%
\subsection{Optimality conditions for worst-case \apost performance
with noisy measurements} 
\label{sec:oc apost noise}

To apply the Bisection-Feasibility Method as described in the previous
section, replace $O_i$ with $O_i^\noisy$ everywhere in
\refeq{feas s}. Thus the primal (feasibility) problem becomes,

\beq[eq:feas s noisy]
\bea{ll}
\mbox{minimize} & s
\\
\mbox{subject to}
&
\trace\ O_i^\noisy A_i(\del) \leq s
\\
&
O_i^\noisy = \sum_{j=1}^m\ \nu_{ij} O_j,
\;
i=1,\ldots,m
\\
&
\sum_{i=1}^m\ O_i = I_n,
\;\;\;
O_i \geq 0,\;
i=1,\ldots,m
\eea
\eeq
The Lagrange function is then,
\beq[eq:lag noisy]
L(s,O,\lam,Z,Y)
=
s+
\sum_{j=1}^m\
\lam_j (\trace\ O_j^\noisy A_j(\del)-s)
-\sum_{i=1}^m\ \trace\ Z_i O_i
+
\trace\ Y\Big( I_n - \sum_{i=1}^m\ O_i \Big)
\eeq
with Lagrange multipliers $\lam_i\in\Rbf,\ \lam_i\geq 0$ for the
inequality constraint $\trace\ O_i^\noisy A_i(\del) \leq s$, $Z_i\in\Cbfnn,\
Z_i \geq 0$ for the inequality constraint $O_i\geq 0$, and
$Y\in\Rbf^{n\times n},\ Y=Y^T$ for the equality constraint $\sum_{i=1}^m\
O_i = I_n$. Eliminating the noisy POVM terms gives,
\beq[eq:lag noisy 1]
L(s,O,\lam,Z,Y)
=
\trace\ Y
+
s\left(1-\sum_{i=1}^m \lam_i\right)
+
\sum_{i=1}^m\ \trace\ O_i\left(A_i(\del,\nu)-Z_i-Y \right)
\eeq
with the $A_i(\del,\nu)$ given by \refeq{optwcpostnoise
amat}. Although not shown, the optimality conditions
\refeq{optwcpostnoise cond} can be established by repeating, {\em
mutadis mutandis}, all the steps in the previous section, \ie,
formulate the dual problem, show that strong duality holds, and so on.

%%%%%%%%%%%%%%%%%%%%%%%%%%%%
\newpage
\bibliographystyle{plain}
\bibliography{rpaper,qpaper}
%%%%%%%%%%%%%%%%%%%%%%%%%%%%%%%%%%%%%%%%%%%
%%%%%%%%%%%%%%%%%%%%%%%%%%%%%%%%%%% 

\end{document}